\documentclass[aps,graphicx,6pt,showkeys,showpacs,onecolumn]{revtex4}
\usepackage{amsmath}
\usepackage{amscd}
\usepackage{graphicx}
\usepackage{subfigure}
\usepackage{appendix}

\begin{document}

\title[Short Title]{Transitionless-based shortcuts for the fast and robust generation of $W$ states}

\author{Ye-Hong Chen$^{1}$}
\author{Yan Xia$^{1,}$\footnote{E-mail: xia-208@163.com}}
\author{Jie Song$^{2, }$\footnote{E-mail: jsong@hit.edu.cn}}
\author{Bi-Hua Huang$^{1}$}

\affiliation{$^{1}$Department of Physics, Fuzhou University, Fuzhou
350002, China\\$^{2}$Department of Physics, Harbin Institute of
Technology, Harbin 150001, China}


\begin{abstract}
  We propose a scheme to generate $W$ states based on transitionless-based shortcuts technique in cavity quantum electrodynamics (QED) system.
  In light of quantum Zeno dynamics, we first effectively design a system whose effective Hamiltonian is equivalent to the
  counter-diabatic driving Hamiltonian constructed by transitionless quantum driving, then, realize the $W$
  states'
  generation within this framework.
  For the sake of clearness, we describe two stale schemes for $W$ states' generation via traditional methods:
  the adiabatic dark-state evolution and the quantum Zeno dynamics.
  The comparison among these three schemes shows the shortcut scheme
  is closely related to the other two but better than them. That is, numerical
  investigation demonstrates that the shortcut scheme is
  faster than the adiabatic one, and more robust
  against operational imperfection than the Zeno one. What is more, the present scheme is also
  robust against decoherence caused by spontaneous emission and photon loss.
\end{abstract}
\pacs {03.67. Pp, 03.67. Mn, 03.67. HK} \keywords{$W$ state;
Transitionless quantum driving; Shortcuts to adiabatic passage.}

\maketitle

\small

\section{Introduction}
Quantum entanglement is an intriguing property of composite systems.
The generation of entangled states for two or more particles is not
only fundamental for demonstrating quantum nonlocality
\cite{JSBPhys65,DMGMAHASAZAjp90}, but also useful in quantum
information processing (QIP) \cite{AKEPrl91,NGSMPrl97}. For
three-qubit entanglement, there are two main kinds of entangled
states , the $W$ states \cite{WDGVJICPRra00} and the
Greenberger-Horne-Zeilinger (GHZ) states \cite{DMGMAHASAZAjp90}.
These two kinds of entangled states cannot be converted to each
other by local operations and classical communications. In recent
years, the $W$ states attract more attentions because of its
robustness against qubit loss and advantages in quantum
teleportation \cite{CHBGBCCRJAPWKWPrl93}. So far, lots of
theoretical schemes have been proposed to generate $W$ states in
different systems via different techniques
\cite{NBAPla05,SBZJob05,JSYXHSSJpb07,TBCTJvZLLESGSAPrl09,XWWGJYYHSMXQip09,RXCLTSPla11,MLYXJSNBAOsa13,YHCYXJSQip13}.
There are two techniques famous for their robustness against
decoherence in proper conditions and have been widely used in QIP:
one is named stimulated Raman scattering involving adiabatic passage
(STIRAP) including their variants
\cite{MPFBWSKBAjp97,KBHTBWSRmp98,NVVTHBWSKBArpc01,PKITMSRmp07,LBCWYLpl14}, and
the other one is Quantum Zeno dynamics (QZD)
\cite{BMECGSJmp77,WMIDJHJJBDJWPra90,PKHWTHAZMAKPrl95,PFSPPrl02,PFVGGMSPECGSPla00}.
Generally speaking, the adiabatic passage technique is robust
against variations in the experimental parameters and atomic
spontaneous emission. To restrain the influence of photon leakage on
the fidelity, a widely used way is choosing parameters to reduce
populations of the intermediate excited states. However, such
operation inevitably increase the operation time. As is known to
all, using adiabatic technique (we name it ``adiabatic scheme'' for
short in this paper) asks for an adiabatic condition that the change
of a system's Hamiltonian in time is managed to be slow to make sure
each of the eigenstates of the system evolves along itself. Using
QZD method, by contrast, might be faster than using adiabatic
passage. But that depends, especially in multiparticle systems.
Usually, in a scheme based on QZD (we name it ``Zeno scheme'' for
short in this paper), we consider the system's Hamiltonian as
$H=H_{obs}+KH_{meas}$, where $H_{obs}$ is the Hamiltonian of the
quantum system investigated, $K$ is the coupling constant, and
$H_{meas}$ is viewed as an additional interaction Hamiltonian
performing the measurement. When $K\rightarrow\infty$, the system's
effective Hamiltonian is approximated as
$H_{Z}=\sum_{n}(K\xi_{n}P_{n}+P_{n}H_{obs}P_{n})$, where $P_{n}$ is
the $n$th eigenvalue projection of $H_{meas}$ with eigenvalue
$\xi_{n}$. Similar to the adiabatic passage, there is also a limited
condition in a Zeno scheme that limits the system's speed: the Zeno
condition $K\rightarrow\infty$. It has been confirmed by lots of
schemes that using QZD for QIP is usually robust against photon
leakage but sensitive to the atomic spontaneous emission. Therefore,
in order to restrain the influence of atomic spontaneous emission,
some researchers introduced detunings between the atomic transitions
to decrease the population of atomic excited states. That also
inevitably increases the operation time. In addition, the operation
time required in a scheme via QZD always needs to be controlled
accurately, which increases the difficulty to realize the scheme in
experiment. As we know, the operation time for a method is the
shorter the better, otherwise, the method may be useless because the
dissipation caused by decoherence, noise, and losses on the target
state increases with the increasing of the interaction time. Many
experiments also desire fast and robust theoretical methods because
high repetition rates contribute to the achievement of better
signal-to-noise ratios and better accuracy.

Therefore, fast and noise-resistant generation of entangled states
becomes a research hotspot in recent years, especially, after the
technique named ``Shortcuts to adiabatic passage'' (STAP)
\cite{XCILARDGOJGMPra10,ETSISMGMMACDGOARXCJGMAmop13} was proposed.
This technique is related to adiabatic passage but successfully
breaks the limit of the adiabatic condition. It describes a fast
adiabatic-like process which is not really adiabatic but leading to
the same final populations with adiabatic process. Newly, STAP has
shown its charm in theory and experiment
\cite{AdCPrl13,XCARSSADCDDOJGMPrl10,SMKNPrspca10Pra11,YHCYXQQCJSPra14,MLYXLTSJSLp14,YHCYXQQCJSLpl14,MLYXLTSJSNBAPra14,ARXCDAJGMNjp12,
JFSXLSPVGLPra10,JFSXLSPCPVGLEpl11,AWFZTRSTDOKTMHKSFSKUPPrl12,SYTXCOl12,JGMXCARDGOJpb09,XCJGMPra10,JFSPCGLPVNjp11,
ETSIXCARDGOJGMPra11,XCETDSJSLJGMPra11,ETXCMMSSARJGMNjp12,YLLAWZDWPra11,AdCPra11,YHCYXQQCJSarXiv142,Prl109100403,
Pra89053408,Njp16015025,Pra90060301,Pra08743402,Pra89043408,Pra8705250289063412,XCETJGMPra10}.
In 2014, by using transitionless tracking algorithm under large
detuning condition , Lu \emph{et al.} proposed an effective scheme
to implement fast populations transfer and fast maximum entanglement
preparation between two atoms in a cavity \cite{MLYXLTSJSNBAPra14}.
The idea inspires that using some traditional methods to approximate
a complicated Hamiltonian into an effective and simple one first,
then constructing shortcuts for the effective Hamiltonian might be a
promising method to speed up a system. Soon after that, Chen
\emph{et al.} first combined invariant-based inverse engineering
with Zeno subspaces to construct shortcuts to perform fast and
noise-resistant populations transfer for multiparticle systems
\cite{YHCYXQQCJSPra14}. In their method, they demonstrated that
besides constructing STAP, slightly broking down the Zeno condition
under certain conditions is another simple way to speed up the
evolution. Soon after that, similar ideas with slightly breaking the
Zeno condition down are rapidly used to perform fast and
noise-resistant QIP
\cite{YHCYXQQCJSLpl14,YHCYXQQCJSarXiv142,YLSLSQCWXJSZarxiv14,YLXJarxiv14,LCSYXJSJmo14}.

Motivated by the above analysis, we discuss how to construct STAP to rapidly generate $W$ states
by using the approach of ``transitionless tracking algorithm'' in cavity QED systems.
Different from ref. \cite{YHCYXQQCJSLpl14} which proposed a method
through combining Lewis-Riesenfeld theory and Zeno subspaces to generate
a $N$-atom $W$ state by $N+1$ atoms, we do not need to abandon any atoms.
An $N$-atom $W$ is fast generated directly by $N$ atoms in one step.
In order to explain the charm by using STAP to generate $W$ states,
we first give a brief description about generation of $W$ states via two
traditional methods (STIRAP and QZD). Then, we propose the scheme by using transitionless tracking algorithm in detail.
The comparison among these three schemes demonstrates that the shortcut scheme is not only
faster than the adiabatic one, and more robust against operational imperfection than the Zeno one.
What is more, this method might be promising when it comes to generation of multi-level and multi-qubit entangled states,
i.e., the singlet states.

The rest of the paper is organized as follows. In section \textrm{II}, we give the model of atom-cavity system and
introduce two schemes to generate $W$ states via traditional adiabatic method and Zeno method.
Then in section \textrm{III}, we use the transitionless-based shortcuts method
to propose a fast and noise-resistant scheme to generate $W$ states. The conclusion is derived in section \textrm{IV}.

\begin{figure}
 \scalebox{0.28}{\includegraphics {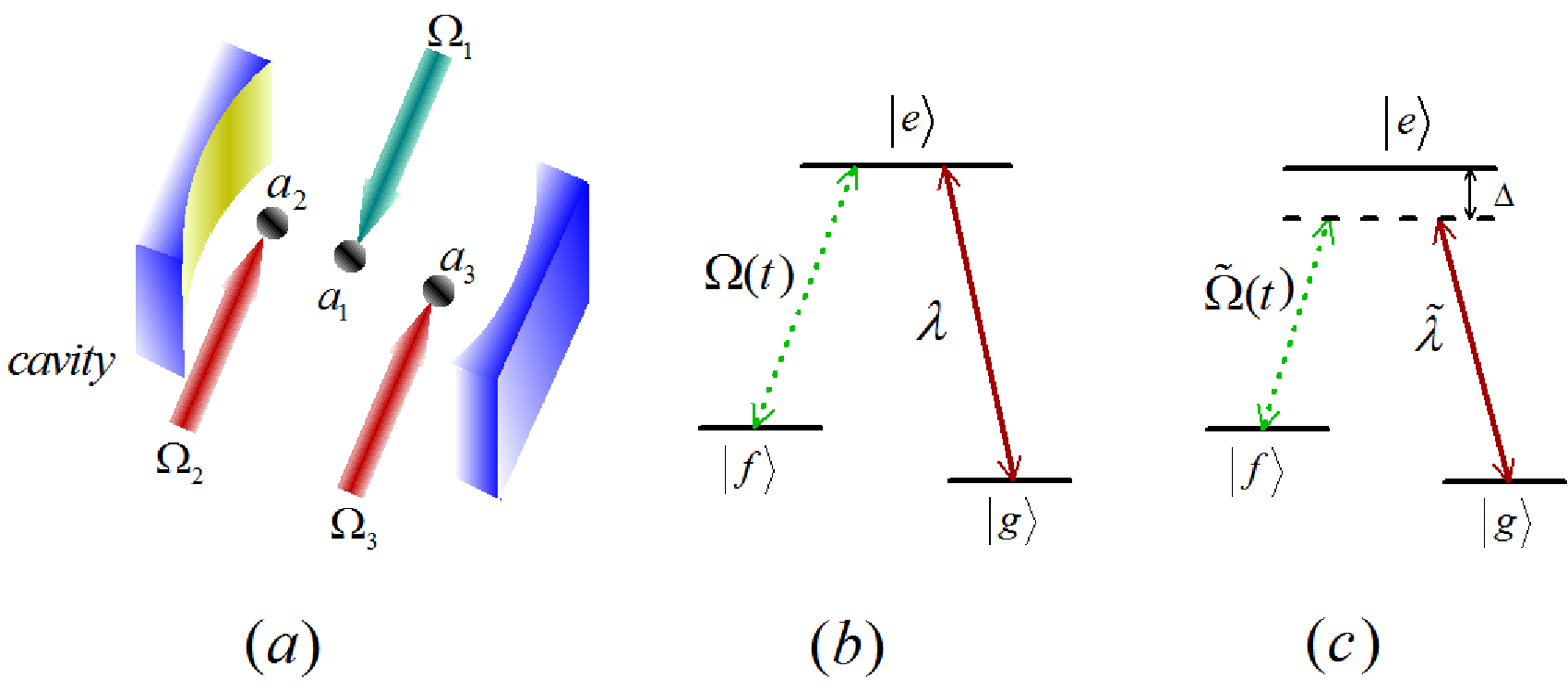}}
 \caption{(a) The experimental set-up diagram.
          (b) The atomic level configuration for traditional schemes (STIRAP and QZD).
          (c) The atomic level configuration for APF system.}
 \label{model}
\end{figure}

\section{theoretical Generation of $W$ states in a three-atom system}
For simplicity, we assume that three $\Lambda$-type atoms ($a_{1}$, $a_{2}$, $a_{3}$) are trapped in a cavity ($c$) as shown in Figs. \ref{model} (a) and (b),
each atom has an excited state $|e\rangle$ and two ground states $|f\rangle$ and $|g\rangle$.
Atomic transition $|f\rangle\leftrightarrow|e\rangle$ is resonantly driven by classical field $\Omega(t)$,
and the transition $|g\rangle\leftrightarrow|e\rangle$ is coupled resonantly
to the cavity with coupling $\lambda$. Under the rotating-wave approximation (RWA), the
interaction Hamiltonian for this system reads
\begin{eqnarray}\label{eq1-1}
  H_{I}=\sum_{k=1}^{3}{\Omega_{k}(t)|e\rangle_{k}\langle f|+\lambda_{k}a|e\rangle_{k}\langle g|+H.c.},
\end{eqnarray}
where $a$ is the annihilation operator of the cavity. 
We assume the initial state of the system is $|f,g,g\rangle_{1,2,3}|0\rangle_{c}$, the system
will evolve within a single-excitation subspace $\forall$ spanned by:
\begin{eqnarray}\label{eq1-2}
  |\psi_{1}\rangle&=&|f,g,g\rangle_{1,2,3}|0\rangle_{c},   \cr
  |\psi_{2}\rangle&=&|e,g,g\rangle_{1,2,3}|0\rangle_{c},   \cr
  |\psi_{3}\rangle&=&|g,g,g\rangle_{1,2,3}|1\rangle_{c},   \cr
  |\psi_{4}\rangle&=&|g,e,g\rangle_{1,2,3}|0\rangle_{c},   \cr
  |\psi_{5}\rangle&=&|g,f,g\rangle_{1,2,3}|0\rangle_{c},   \cr
  |\psi_{6}\rangle&=&|g,g,e\rangle_{1,2,3}|0\rangle_{c},   \cr
  |\psi_{7}\rangle&=&|g,g,f\rangle_{1,2,3}|0\rangle_{c}.
\end{eqnarray}
Here it is worth noting that in a natural case, the atoms are usually in the same state initially, i.e., the steady state $|g\rangle$.
So, it is necessary to prepare the initial state $|\psi_{1}\rangle$ before implementing the scheme. That is, a population transfer 
$|g\rangle_{1}\rightarrow|f\rangle_{1}$ is imperative before the scheme. Fortunately, such operation is not hard to be realized.
$\pi$ pulse, stimulated Raman adiabatic passage, large detuning dynamics, and many other techniques are applicable to transfer population from $|g\rangle_{1}$ to $|f\rangle_{1}$.
Therefore, in subspace $\forall$, the interaction Hamiltonian is simplified as (we set $\lambda_{k}=\lambda$ to be constant coupling coefficients)
\begin{eqnarray}\label{eq1-3}
  H_{0}&=&H_{al}+H_{ac},                           \cr
  H_{al}&=&\Omega_{1}|\psi_{2}\rangle\langle\psi_{1}|+\Omega_{2}|\psi_{4}\rangle\langle\psi_{5}|+\Omega_{3}|\psi_{6}\rangle\langle\psi_{7}|+H.c.,     \cr
  H_{ac}&=&\lambda(|\psi_{2}\rangle+|\psi_{4}\rangle+|\psi_{6}\rangle)\langle\psi_{3}|+H.c..
\end{eqnarray}
For the sake of clearness, we will describe two traditional
different methods (STIRAP and QZD) to generate $W$ states in brief.

\begin{figure}
 \scalebox{0.18}{\includegraphics {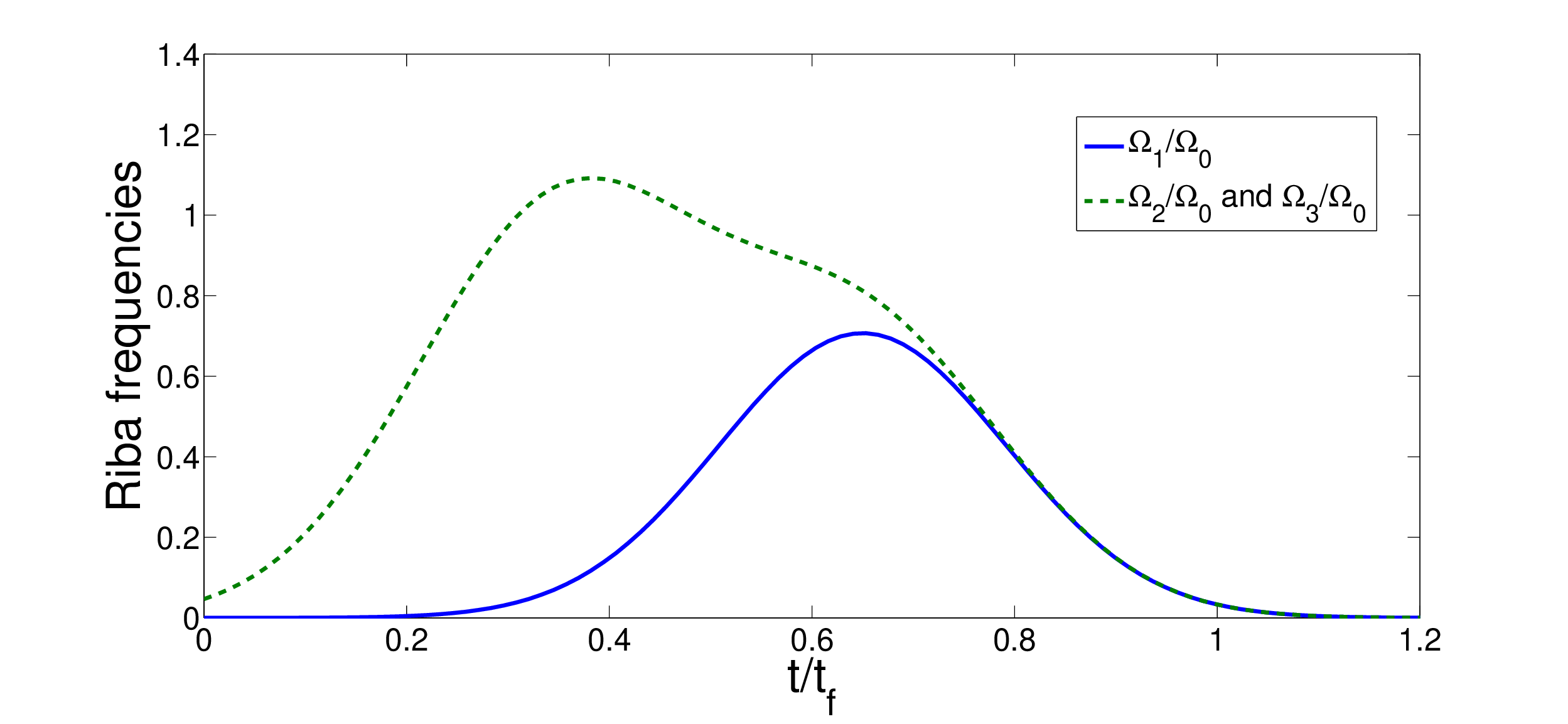}}
 \caption{Dependence on $t/t_{f}$ of $\Omega_{1}/\Omega_{0}$ and $\Omega_{s}/\Omega_{0}$.}
 \label{O1O2}
 \scalebox{0.2}{\includegraphics {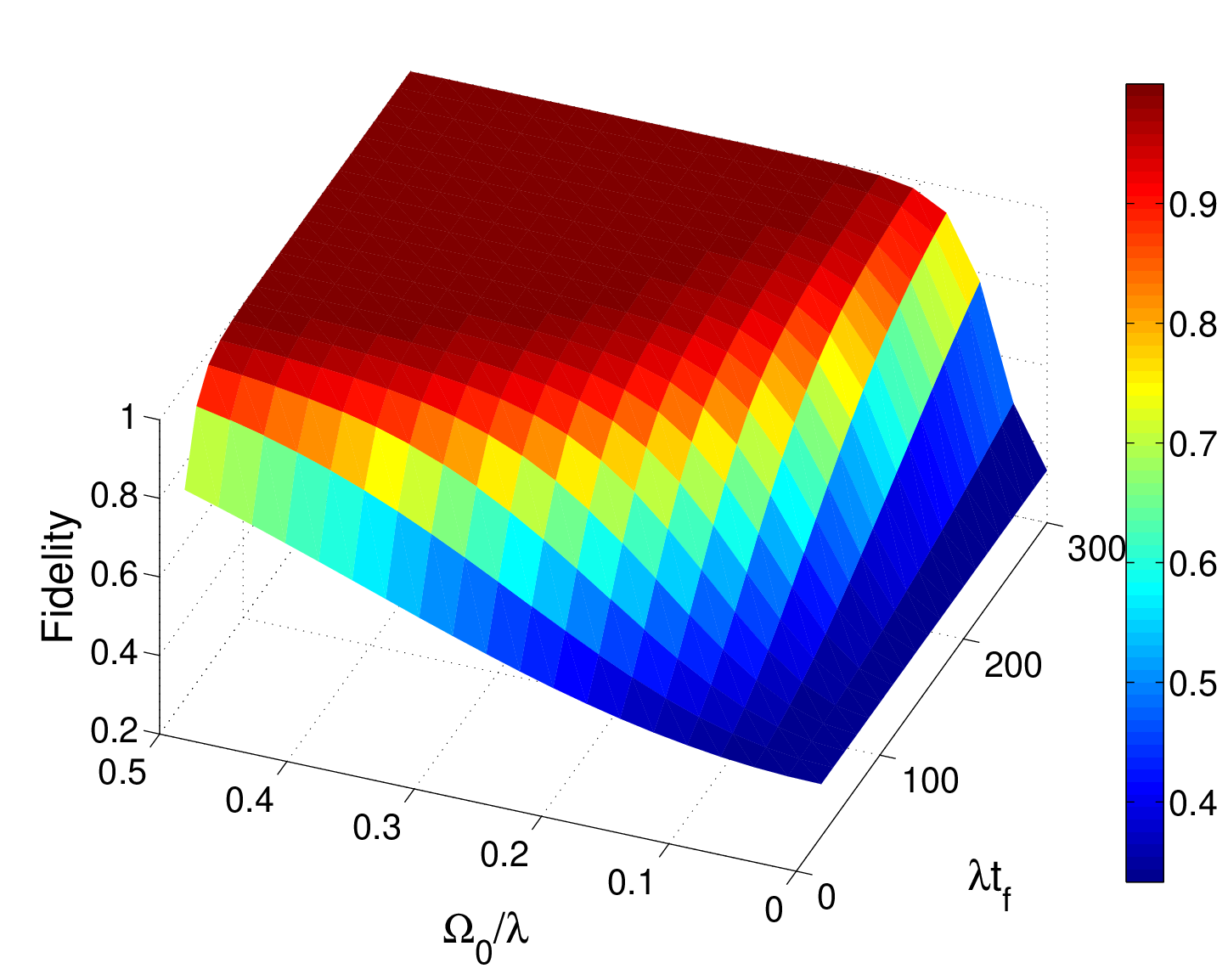}}
 \caption{The fidelity of the $W$ state generated via STIRAP versus $\Omega_{0}/\lambda$ and $\lambda t_{f}$.}
 \label{Adicon}
\end{figure}

\begin{figure}
 \renewcommand\figurename{\small FIG.}
 \centering \vspace*{8pt} \setlength{\baselineskip}{10pt}
 \subfigure[]{
 \includegraphics[scale = 0.18]{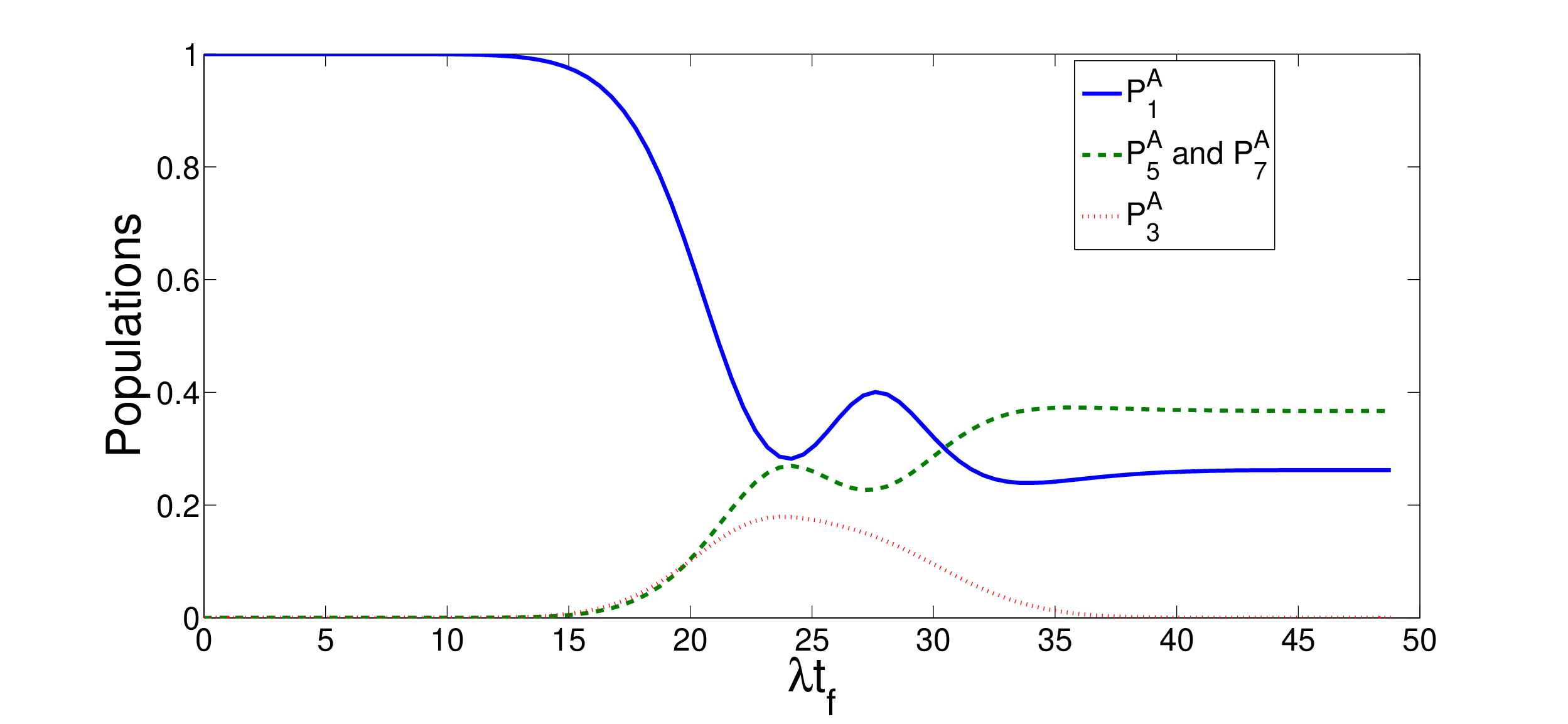}}
 \subfigure[]{
 \includegraphics[scale = 0.18]{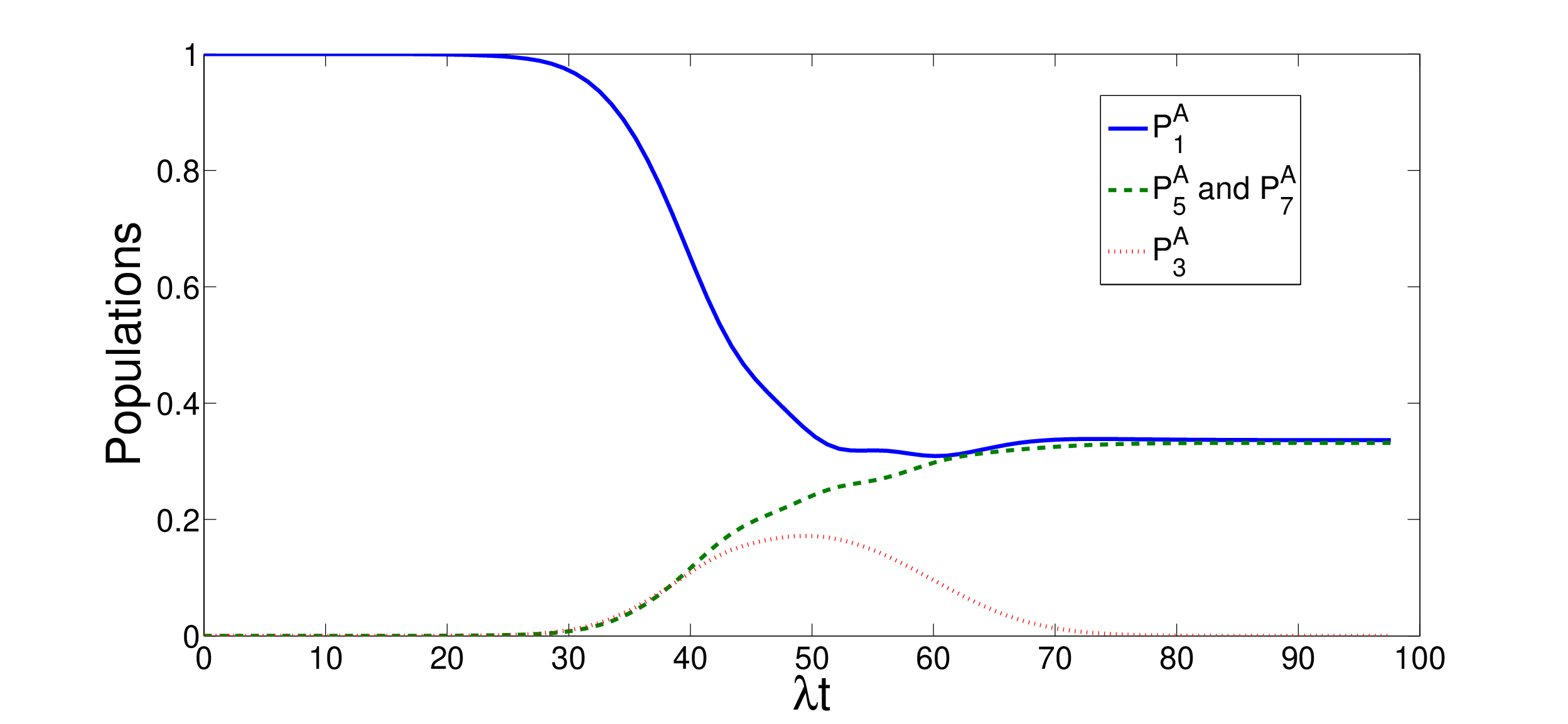}}
 \subfigure[]{
 \includegraphics[scale = 0.18]{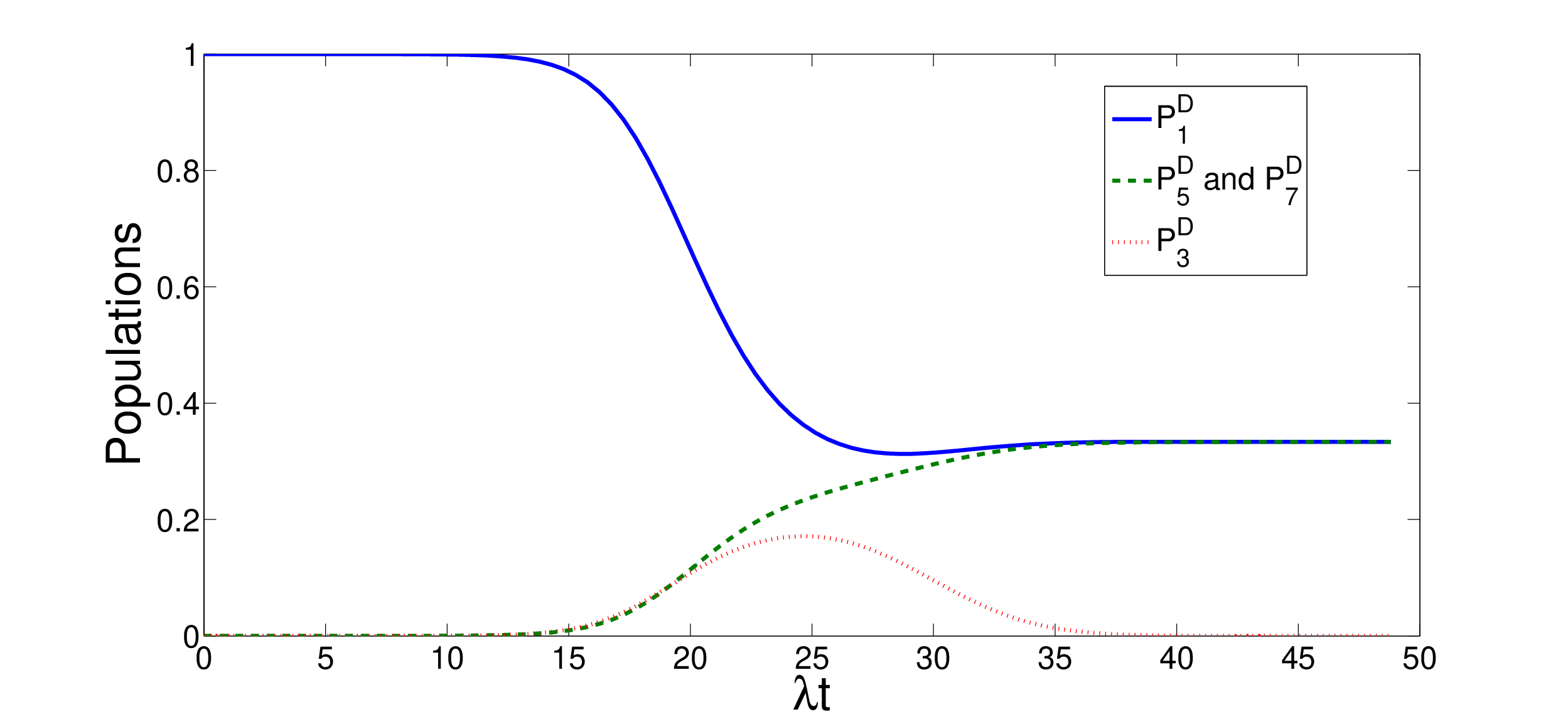}}
 \caption{
        The time evolution of the populations for
        states $|\psi_{1}\rangle$, $|\psi_{3}\rangle$, $|\psi_{5}\rangle$, and $|\psi_{7}\rangle$ via STIRAP
        in different cases ((a) and (b) are plotted through relation $P^{A}_{m}=|\langle\psi_{m}|\rho|\psi_{m}\rangle|$, (c) is plotted
        through relation $P^{D}_{m}=|\langle\psi_{m}|\Psi_{0}\rangle|^{2}$ ($m=1,3,5,7$)):
        (a) when $\Omega_{0}=\lambda$ and $t_{f}=40/\lambda$,
        (b) when $\Omega_{0}=\lambda$ and $t_{f}=80/\lambda$,
        (c) when $\Omega_{0}=\lambda$ and $t_{f}=40/\lambda$.
          }
 \label{P1357adi}
\end{figure}

\subsection{Based on STIRAP}
For the Hamiltonian in eq. (\ref{eq1-3}) in
the single-excitation subspace, we easily find a dark state
\begin{eqnarray}\label{eq1a-1}
  |\Psi_{0}(t)\rangle=\frac{1}{\sqrt{N_{D}}}(\frac{|\psi_{1}\rangle}{\Omega_{1}}+\frac{|\psi_{5}\rangle}{\Omega_{2}}+\frac{|\psi_{7}\rangle}{\Omega_{3}}-\frac{|\psi_{3}\rangle}{\lambda}),
\end{eqnarray}
where $N_{D}=(\frac{1}{\Omega_{1}})^{2}+(\frac{1}{\Omega_{2}})^{2}+(\frac{1}{\Omega_{3}})^{2}+(\frac{1}{\lambda})^{2}$ is the normalization coefficient.
If we choose $\Omega_{2},\Omega_{3},\lambda\gg\Omega_{1}$ at first and make sure that the adiabatic condition
$|\langle \Psi_{0}|\partial_{t}\Psi_{n}\rangle|\ll|\xi_{n}|$ is satisfied, where $|\Psi_{n}\rangle$ is the
$n$th eigenstate with nonzero eigenvalue $\xi_{n}$, the initial state
$|\psi_{1}\rangle=|\Psi_{0}(0)\rangle$ will follow $|\Psi_{0}(t)\rangle$ closely. Then, we slowly decrease $\Omega_{2}$ and $\Omega_{3}$ while increase $\Omega_{1}$ until
$\Omega_{1}=\Omega_{2}=\Omega_{3}\ll\lambda$ at time $t_{f}$. Accordingly, the dark state becomes
$|\Psi_{0}(t_{f})\rangle=\frac{1}{\sqrt{3}}(|\psi_{1}\rangle+|\psi_{5}\rangle+|\psi_{7}\rangle)$ which is the $W$ state.
As shown in Fig. \ref{O1O2}, to complete this process, we choose the Rabi frequencies as (we set $\Omega_{2}=\Omega_{3}=\Omega_{s}$)
\begin{eqnarray}\label{eq1a-2}
   \Omega_{1}&=&\sin{\alpha}\Omega_{0}\exp[\frac{-(t-t_{0}-t_{f}/2)^{2}}{t_{c}^{2}}], \cr\cr
   \Omega_{s}&=&\Omega_{0}\exp[\frac{-(t+t_{0}-t_{f}/2)^{2}}{t_{c}^{2}}]              \cr\cr
                  &&+\cos{\alpha}\Omega_{0}\exp[\frac{-(t-t_{0}-t_{f}/2)^{2}}{t_{c}^{2}}],
\end{eqnarray}
where $\Omega_{0}$ is the amplitude and $\{t_{0},t_{c}\}$ are
related parameters. To meet the conditions mentioned above, we
choose $\tan{\alpha}=1$, $t_{0}=0.15t_{f}$, and $t_{c}=0.2t_{f}$.
Generally speaking, the adiabatic condition is satisfied better with
a relatively large $\Omega_{0}$ because the nonzero eigenvalue
$\xi_{n}$ is proportional to $\Omega_{0}$. Fig. \ref{Adicon} shows
the fidelity of the $W$ state in adiabatic scheme versus the
interaction time $t_{f}$ and $\Omega_{0}$. The fidelity $F$ for any
target state $|\psi\rangle$ is given through the relation
$F=|\langle\psi|\rho|\psi\rangle|$, where $\rho$ is the density
operator given through $\dot{\rho}=i[\rho,H]$. As shown in Fig.
\ref{Adicon}, the fidelity is getting higher with both the increases
of $\Omega_{0}$ and $t_{f}$. It seems that when $\Omega_{0}\times
t_{f}\geq40$, a high-fidelity $W$ state is achievable. That means
when $\Omega_{0}$ is large enough, we also can create a $W$ state in
a short interaction time via adiabatic passage. But further
investigation tells us that the system's evolution is far different
from adiabatic dark-state evolution with a relatively large
$\Omega_{0}$ and a short interaction time $t_{f}$. Fig.
\ref{P1357adi} (a) shows the time-dependent populations for states
\{$|\psi_{1}\rangle$, $|\psi_{5}\rangle$, $|\psi_{7}\rangle$,
$|\psi_{3}\rangle$\} when $\Omega_{0}=\lambda$ and
$t_{f}=40/\lambda$, Fig. \ref{P1357adi} (b) shows the time-dependent
populations for states \{$|\psi_{1}\rangle$, $|\psi_{5}\rangle$,
$|\psi_{7}\rangle$, $|\psi_{3}\rangle$\} when $\Omega_{0}=\lambda$
and $t_{f}=80/\lambda$, and Fig. \ref{P1357adi} (c) shows
time-dependent evolution of the dark state $|\Psi_{0}(t)\rangle$.
The comparison of these three figures draws a result that even with
a large $\Omega_{0}$, a relatively long interaction time is still
necessary to make sure the controlling parameters change slowly
enough to allow adiabatic passage from an initial state to a target
state. In addition, because a relatively large $\Omega_{0}$  might
cause that the RWA is no longer effective for the system, and it
also means a great population of state $|\psi_{3}\rangle$ including
a cavity-excited state that makes the system sensitive to the cavity
photon leakage, it is better to choose a relatively small
$\Omega_{0}$ and a long interaction time $t_{f}$ for an adiabatic
process.

\begin{figure}
 \scalebox{0.18}{\includegraphics {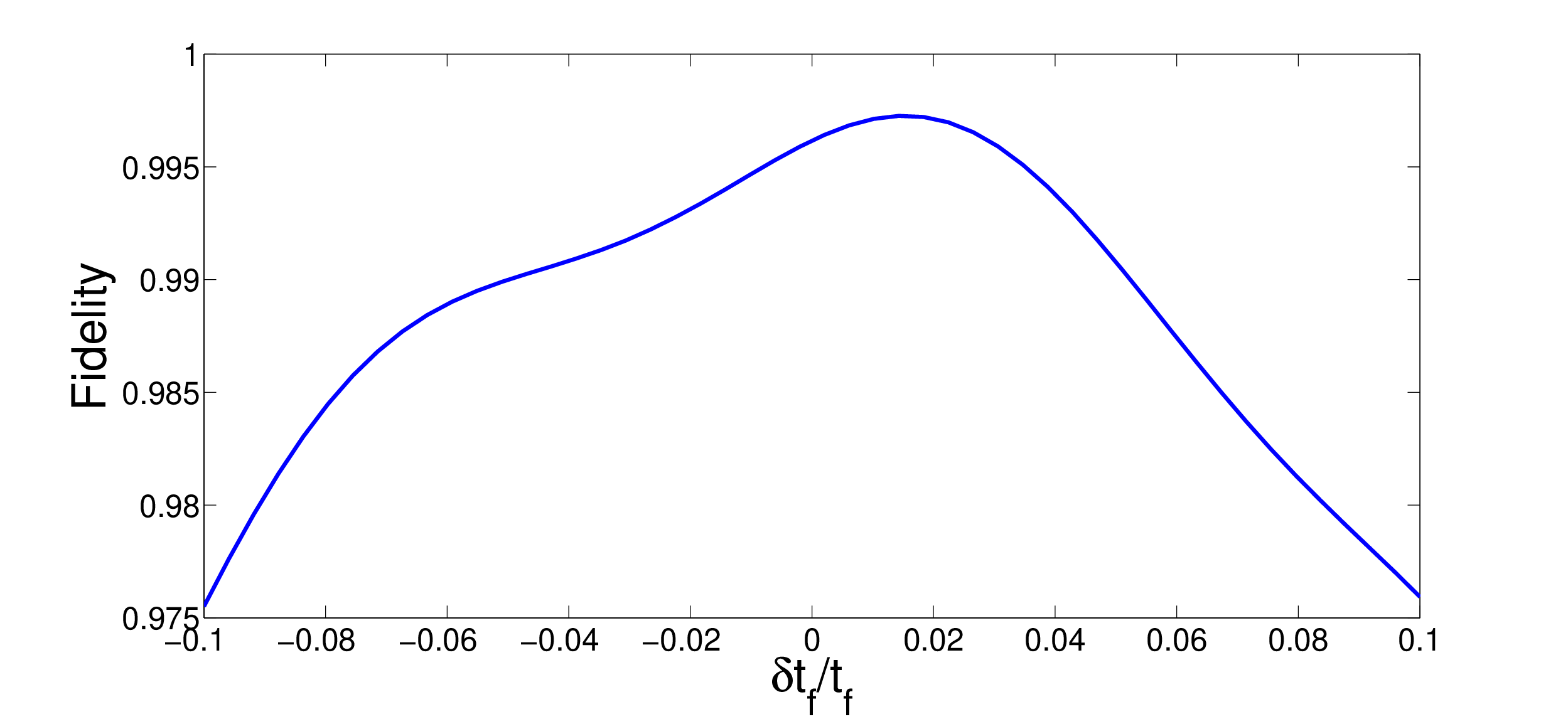}}
 \caption{The fidelity of the $W$ state  generated via QZD versus the variations of $t_{f}$.}
 \label{ZenodeltaT}
\end{figure}

\subsection{Based on QZD}
Before we start using QZD to create a three-atom $W$ state, we set $\Omega_{2}=\Omega_{3}=\Omega_{s}$ and use two orthogonal vectors
$|\mu_{+}\rangle=\frac{1}{\sqrt{2}}(|\psi_{4}\rangle+|\psi_{6}\rangle)$ and $|\mu_{-}\rangle=\frac{1}{\sqrt{2}}(|\psi_{4}\rangle-|\psi_{6}\rangle)$ to
rewrite the Hamiltonian in eq. (\ref{eq1-3}) as
\begin{eqnarray}\label{eq1b-1}
  H_{al}&=&\Omega_{1}|\psi_{2}\rangle\langle\psi_{1}|+\frac{\Omega_{s}}{\sqrt{2}}|\mu_{+}\rangle(\langle\psi_{5}|+\langle\psi_{7}|)             \cr\cr
                                                      &&+\frac{\Omega_{s}}{\sqrt{2}}|\mu_{-}\rangle(\langle\psi_{5}|-\langle\psi_{7}|)+H.c.,  \cr\cr
  H_{ac}&=&\lambda(|\psi_{2}\rangle+\sqrt{2}|\mu_{+}\rangle)\langle\psi_{3}|+H.c..
\end{eqnarray}
It is obvious that when the initial state is $|\psi_{1}\rangle$, the terms containing $|\mu_{-}\rangle$ are negligible
because they are decoupled to the time evolution of initial state.
Then, under the condition $\Omega_{1},\Omega_{s}\ll\sqrt{3}\lambda$ (Zeno condition), the subspace $\forall$  is
split into three Zeno subspaces according to the degeneracy
of eigenvalues of $H_{ac}$,
\begin{eqnarray}\label{eq1b-2}
  Z_{0}&=&\{|\psi_{1}\rangle,|\psi_{5}\rangle,|\psi_{7}\rangle,|\phi_{1}\rangle\},  \cr
  Z_{+}&=&\{|\phi_{2}\rangle\}, \ \ Z_{-}=\{|\phi_{3}\rangle\}.
\end{eqnarray}
where
\begin{eqnarray}\label{eq1b-3}
  |\phi_{1}\rangle&=&\frac{1}{\sqrt{3}}(-\sqrt{2}|\psi_{2}\rangle+|\mu_{+}\rangle),  \cr
  |\phi_{2}\rangle&=&\frac{1}{\sqrt{6}}(|\psi_{2}\rangle+\sqrt{3}|\psi_{3}\rangle+\sqrt{2}|\mu_{+}\rangle),  \cr
  |\phi_{3}\rangle&=&\frac{1}{\sqrt{6}}(|\psi_{2}\rangle-\sqrt{3}|\psi_{3}\rangle+\sqrt{2}|\mu_{+}\rangle),
\end{eqnarray}
corresponding eigenvalues $\varepsilon_{1}=0$, $\varepsilon_{2}=\sqrt{3}\lambda$, and $\varepsilon_{3}=-\sqrt{3}\lambda$.
Under the Zeno condition, we obtain the effective Hamiltonian governing the evolution
\begin{eqnarray}\label{eq1b-4}
  H_{Z}=-\frac{\sqrt{2}\Omega_{1}}{\sqrt{3}}|\phi_{1}\rangle\langle\psi_{1}|
        +\frac{\Omega_{s}}{\sqrt{3}}|\phi_{1}\rangle\langle\zeta|+H.c.,
\end{eqnarray}
where $|\zeta\rangle=\frac{1}{\sqrt{2}}(|\psi_{5}\rangle+\psi_{7}\rangle)$.
When $\Omega_{1}$ and $\Omega_{s}$ are constant parameters,
the general evolution of eq. (9) by solving the Schr\"{o}dinger equation $i\partial_{t}|\psi(t)\rangle=H_{Z}|\psi(t)\rangle$ in time $t$ is
\begin{eqnarray}\label{eq1b-5}
  |\psi(t)\rangle&=&\frac{\Omega_{s}^{2}+2\Omega_{1}^{2}\cos{\beta t}}{3\beta^{2}}|\psi_{1}\rangle
                    +\frac{i\Omega_{1}\sin{\beta t}}{\sqrt{3}\beta}|\phi_{1}\rangle                               \cr\cr
                    &&+\frac{\sqrt{2}\Omega_{1}\Omega_{s}-\sqrt{2}\Omega_{1}\Omega_{s}\cos{\beta t}}{3\beta^{2}}|\zeta\rangle,
\end{eqnarray}
where $\beta=\sqrt{\frac{2\Omega_{1}^{2}+\Omega_{s}^{2}}{3}}$. By choosing $\Omega_{s}=(1\pm\sqrt{3})\Omega_{1}$ and $t=t_{f}=\frac{\pi}{\beta}$,
the final state becomes $|\psi(t_{f})\rangle=\frac{1}{\sqrt{3}}(|\psi_{1}\rangle+|\psi_{5}\rangle+|\psi_{7}\rangle)$ which is the $W$ state.

In general, the interaction time required in a Zeno scheme is
shorter than that in an adiabatic scheme. For example, in the
present scheme, when we choose relatively large laser pulses, i.e.,
$\Omega_{1}=0.05\lambda$, the interaction time in the Zeno scheme is
only about $t_{f}=\pi/\delta\approx 35.4/\lambda$. However, it is
well known that the interaction time should be controlled accurately
in a scheme via QZD. We plot the fidelity of the $W$ state in Zeno
scheme versus the variation in $t_{f}$ in Fig. \ref{ZenodeltaT}.
Here we define $\delta x=x'-x$ as the deviation of any parameter
$x$, where $x'$ is the actual value and $x$ is the ideal value. It
is clear that a deviation $|\delta t_{f}/t_{f}|=10\%$ causes a
reduction about $3\%$ in the fidelity, which shows the scheme is
sensitive to the variation of the interaction time. In experiment,
if we choose a related parameter $\lambda=1$GHZ, the required
interaction time is $t_{f}=3.54\times 10^{-8}$s. That means, the
experimental researchers should accurately control the interaction
time to ensure the deviation in $t_{f}$ is less than $|\delta
t_{f}|=3.54\times 10^{-9}$s. That is really a challenge in the
current experimental technology. Moreover, known from eq.
(\ref{eq1b-5}), the intermediate state $|\phi_{1}\rangle$ including
atomic-excited states would be greatly populated during the
evolution if $\Omega_{1}$ is too large, and that might make the
system sensitive to the spontaneous emission.

\begin{figure}
 \scalebox{0.2}{\includegraphics {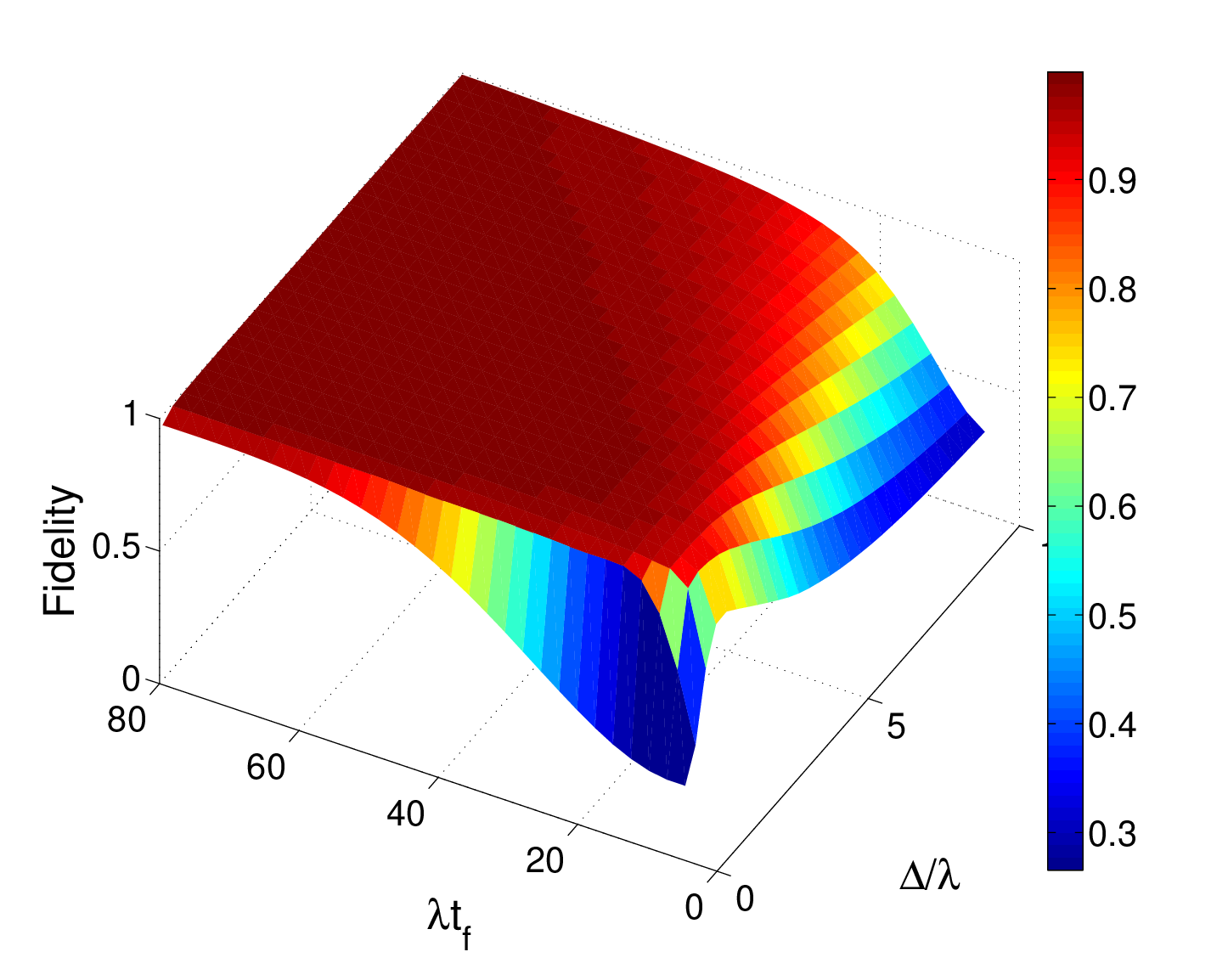}}
 \caption{The fidelity of the $W$ state in the shortcut scheme versus the interaction time $\lambda t_{f}$ and the detuning $\Delta/\lambda$ in the shortcut scheme.}
 \label{Deltatf}
\end{figure}

\section{Using STAP to fast generate a $W$ state}
\begin{figure}
 \renewcommand\figurename{\small FIG.}
 \centering \vspace*{8pt} \setlength{\baselineskip}{10pt}
 \subfigure[]{
 \includegraphics[scale = 0.18]{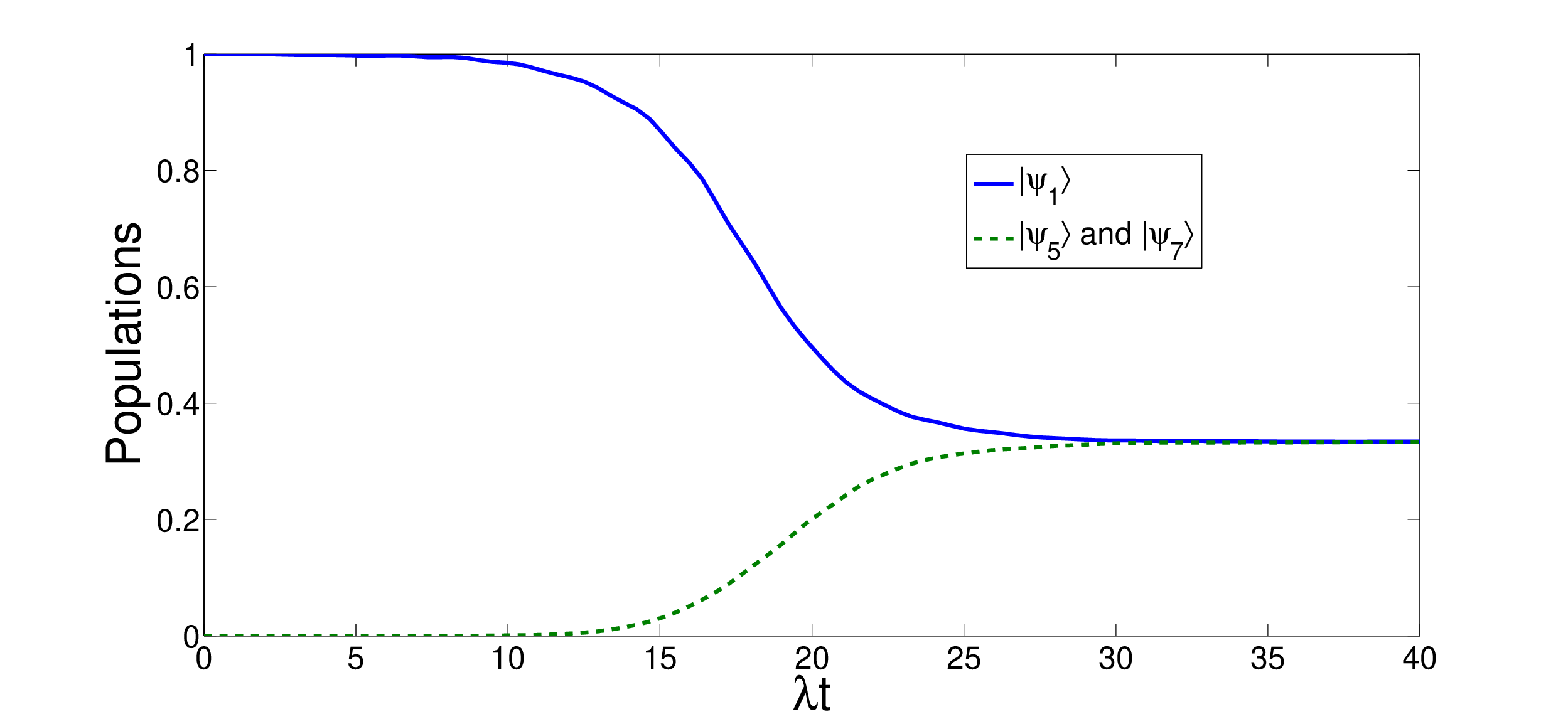}}
 \subfigure[]{
 \includegraphics[scale = 0.18]{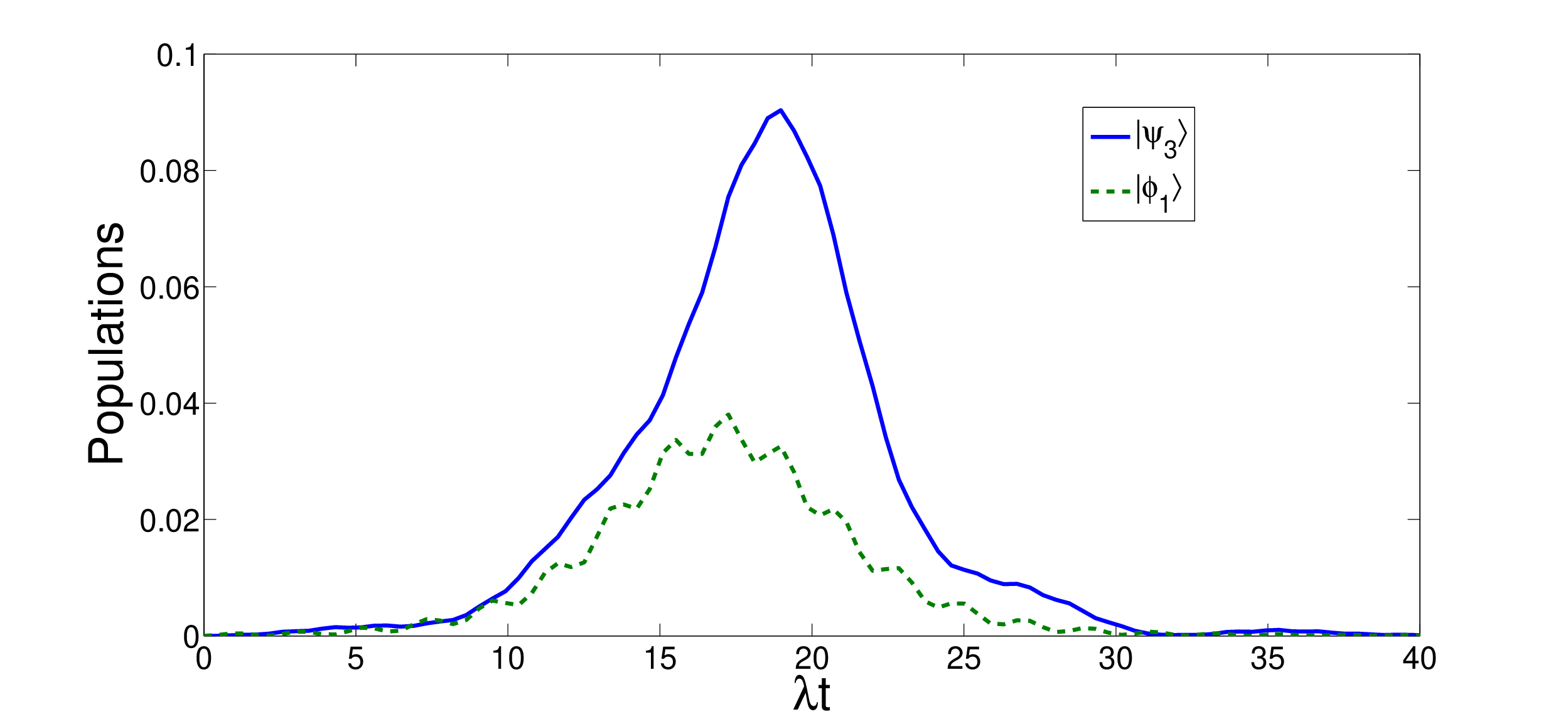}}
 \subfigure[]{
 \includegraphics[scale = 0.18]{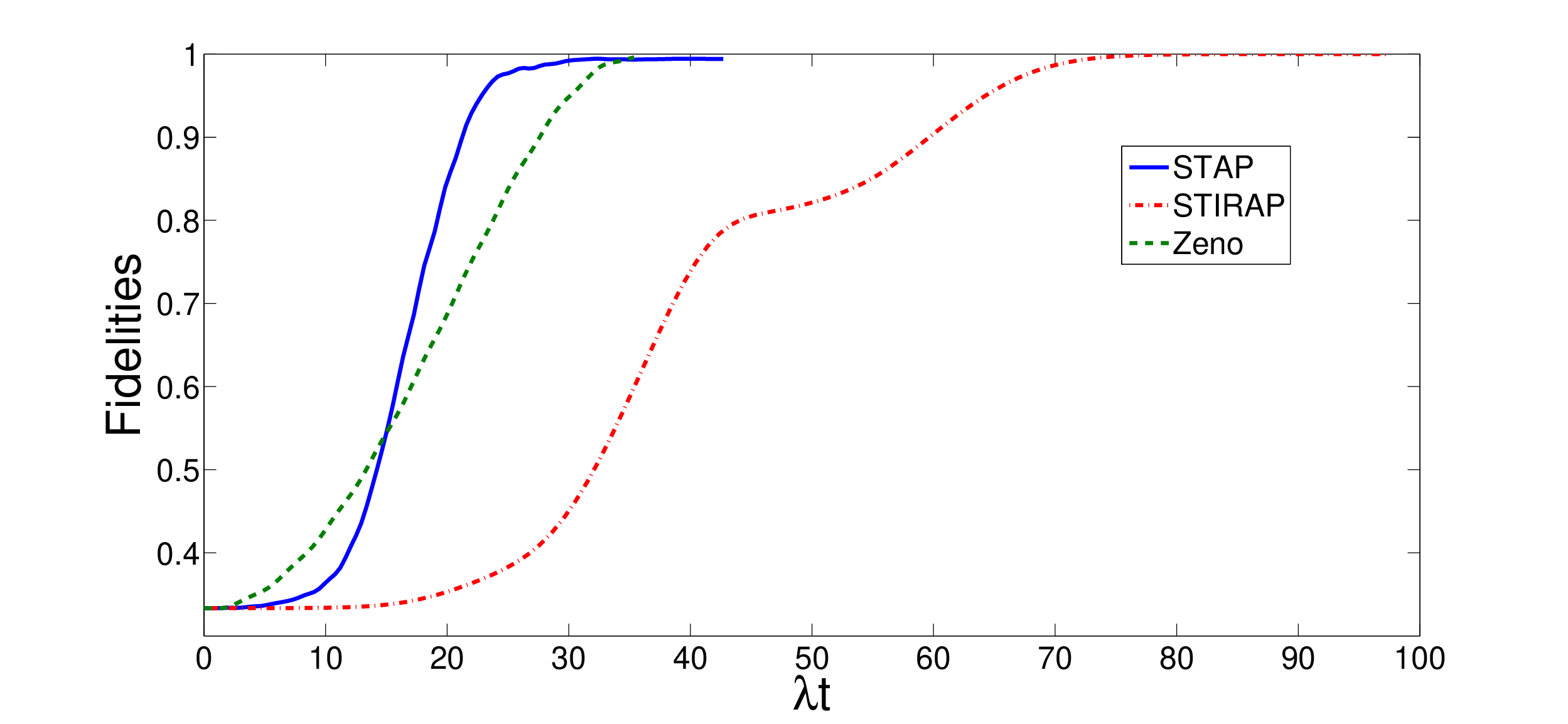}}
 \caption{
        (a) Time evolution of the populations for the states $|\psi_{1}\rangle$, $|\psi_{5}\rangle$ and $|\psi_{7}\rangle$ with \{$t_{f}=35/\lambda$, $\delta=3\lambda$\}.
        (b) Time evolution of the populations for the intermediate states $|\psi_{3}\rangle$ and $|\phi_{1}\rangle$ with \{$t_{f}=35/\lambda$, $\delta=3\lambda$\}.
        (c) The comparison between the fidelities of the three schemes,
            the blue solid curve representing the shortcut scheme is plotted with \{$t_{f}=35/\lambda$, $\delta=3\lambda$\},
            the red dash-dot curve representing the adiabatic scheme is plotted with \{$t_{f}=80/\lambda$, $\Omega_{0}=\lambda$\},
            and the green dash curve representing the Zeno scheme is plotted with \{$t_{f}\approx 35.4/\lambda$, $\Omega_{1}=0.05\lambda$\}.
          }
 \label{Ft}
\end{figure}

\begin{figure}
 \scalebox{0.2}{\includegraphics {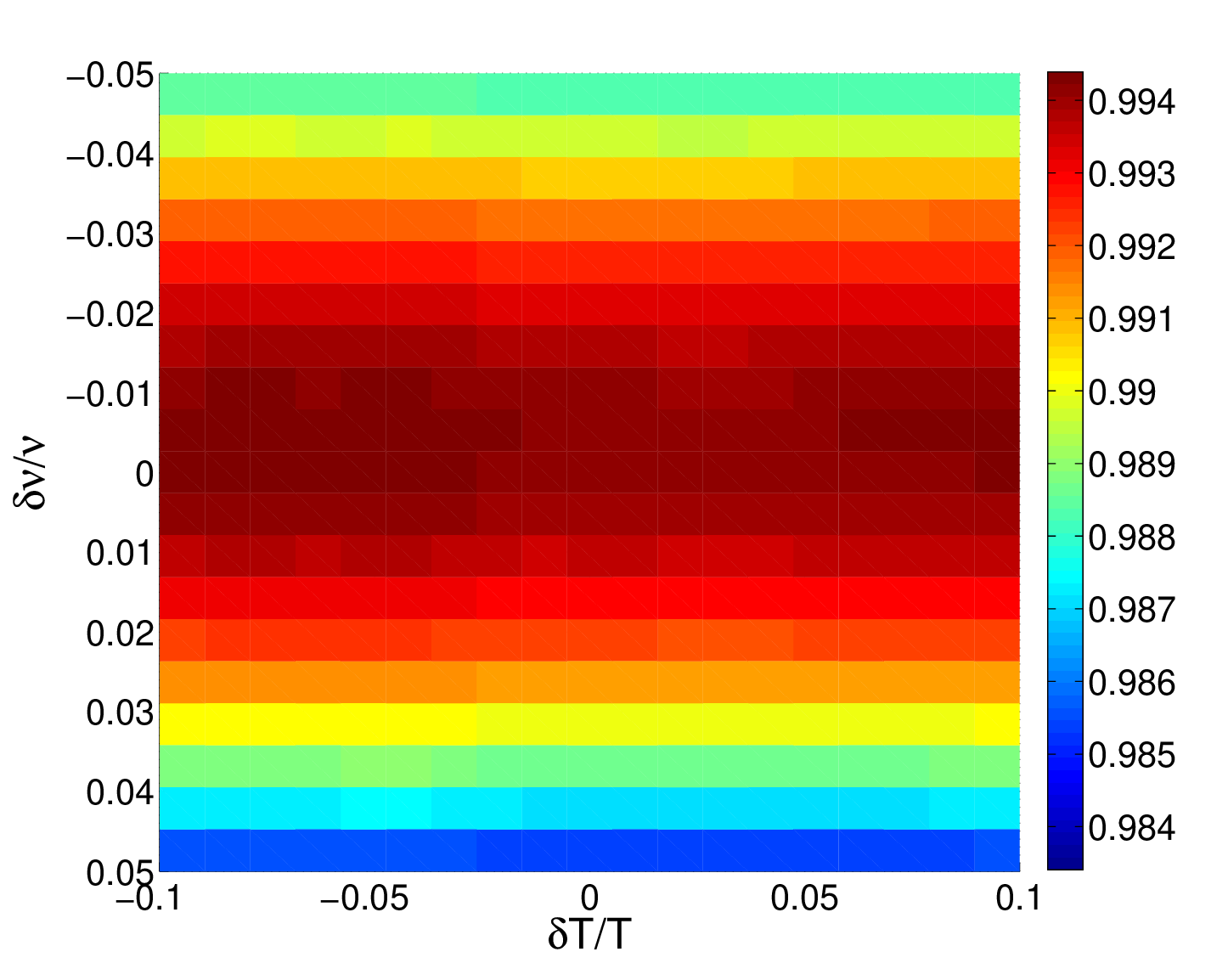}}
 \caption{The fidelity of the $W$ state via STAP versus the variations of $T$ and $\nu$.}
 \label{deltaTZ}
\end{figure}

Different from the two methods (adiabatic scheme and Zeno scheme)
mentioned above, we start from finding a Hamiltonian $H(t)$ which is
related to $H_{0}(t)$ to fast generate a $W$ state via STAP. The key
point to construct shortcuts for a system governed by $H_{0}(t)$ is
to find out a Hamiltonian $H(t)$ which drives the instantaneous
eigenstates $\{|\Psi_{m}\rangle\}$ ($m=0,1,2,3,4$) of $H_{0}(t)$
exactly. Known from Berry's general transitionless tracking
algorithm \cite{MVBJpa09}, the Hamiltonian $H(t)$ can be reverse
engineered from $H_{0}(t)$. And disregarding the effect of phases,
the simplest Hamiltonian $H(t)$ is derived in form of
\begin{eqnarray}\label{eq2-1}
  H(t)=i\sum_{m=0}^{4}{|\partial_{t}\Psi_{m}\rangle\langle\Psi_{m}|}.
\end{eqnarray}
However, it seems impossible to directly design such Hamiltonian from $H_{0}(t)$ according to eq. (\ref{eq1-3}) because the
eigenstates $\{|\Psi_{m}\rangle\}$ given by solving the intrinsic
equation $H_{0}|\Psi_{m}(t)\rangle=E_{m}(t)|\Psi_{m}(t)\rangle$ are very complex such that mathematically
solving eq. (\ref{eq2-1}) seems an outstanding challenge.
So here we make a limiting condition $\Omega_{1},\Omega_{2},\Omega_{3}\ll\lambda$ to simplify the calculation.
Under this condition, by substituting the instantaneous eigenstates $\{|\Psi_{m}\rangle\}$ of $H_{0}$ in eq. (\ref{eq1-3}) into eq. (\ref{eq2-1}),
we obtain
the Hamiltonian that exactly drives the eigenstates of $H_{Z}(t)$
\begin{eqnarray}\label{eq2-5}
  H(t)&=&i\sum_{m=0}^{4}{|\partial_{t}\Psi_{m}\rangle\langle\Psi_{m}|}  \cr\cr
         &=&-i\dot{\theta}|\zeta\rangle\langle\psi_{1}|+H.c.,
\end{eqnarray}
where $\theta=\arctan{\frac{\sqrt{2}\Omega_{1}}{\Omega_{s}}}$, $\Omega_{s}=\Omega_{2}=\Omega_{3}$, and $|\zeta\rangle=\frac{1}{\sqrt{2}}(|\psi_{5}\rangle+|\psi_{7}\rangle)$.
Obviously, there is no way to directly apply a pulse between the state $|\psi_{1}\rangle$ and $|\zeta\rangle$.
So, to realize such a Hamiltonian, we need to find out an alternative physically feasible (APF) system whose
effective Hamiltonian is equivalent to $H(t)$.
The model of the APF system is the same as that in Fig. \ref{model} (a). The difference happens in the atomic level configuration
as shown in Fig. \ref{model} (c), each atom also has three levels $|f\rangle$, $|g\rangle$, and $|e\rangle$.
The transition $|f\rangle\leftrightarrow |e\rangle$ is non-resonantly driven by classical field
with time-dependent Rabi frequency $\tilde{\Omega}(t)$ and detuning $\Delta$.
The transition $|g\rangle\leftrightarrow |e\rangle$ is coupled non-resonantly to the cavity with coupling $\tilde{\lambda}$
and detuning $\Delta$. Similar to the transformation from eq. (\ref{eq1-3}) to eq. (\ref{eq1b-1}),
the interaction Hamiltonian in the subspace $\forall$
(the single-excitation subspace for this model is also spanned by eq. (\ref{eq1-2})) for the present model can be described as
(we set $\tilde{\Omega}_{2}=\tilde{\Omega}_{3}=\tilde{\Omega}_{s}$ and $\tilde{\lambda}_{1}=\tilde{\lambda}_{2}=\tilde{\lambda}_{3}=\lambda$)
\begin{eqnarray}\label{eq2-6}
  \tilde{H}_{0}&=&\tilde{H}_{al}+\tilde{H}_{ac}+\tilde{H}_{e},   \cr\cr
  \tilde{H}_{al}&=&\tilde{\Omega}_{1}|\psi_{2}\rangle\langle\psi_{1}|+\frac{\tilde{\Omega}_{s}}{\sqrt{2}}|\mu_{+}\rangle(\langle\psi_{5}|+\langle\psi_{7}|)  \cr\cr
                 &&+\frac{\tilde{\Omega}_{s}}{\sqrt{2}}|\mu_{-}\rangle(\langle\psi_{5}|-\langle\psi_{7}|)+H.c.,                                              \cr\cr
  \tilde{H}_{ac}&=&H_{ac}=\lambda(|\psi_{2}\rangle+\sqrt{2}|\mu_{+}\rangle)\langle\psi_{3}|+H.c.,                                                           \cr\cr
  \tilde{H}_{e}&=&\Delta|\psi_{2}\rangle\langle\psi_{2}|+\Delta|\mu_{+}\rangle\langle\mu_{+}|+\Delta|\mu_{-}\rangle\langle\mu_{-}|.
\end{eqnarray}
The terms including $|\mu_{-}\rangle$ are also neglected because they are decoupled to
the time evolution of initial state when the initial state is set as $|\psi_{1}\rangle$.
Using the eigenstates of $\tilde{H}_{ac}$ to rewrite this Hamiltonian and
performing the unitary transformation $U=e^{-i\tilde{H}_{ac}t}$, we obtain
\begin{eqnarray}\label{eq2-a}
  \tilde{H}_{al}^{re}&=&{\tilde{\Omega}_{1}}(-\frac{\sqrt{2}}{\sqrt{3}}|\phi_{1}\rangle+\frac{1}{\sqrt{6}}e^{i\epsilon_{2}t}|\phi_{2}\rangle
                                     +\frac{1}{\sqrt{6}}e^{i\epsilon_{3}t}|\phi_{3}\rangle)\langle\psi_{1}| \cr\cr
                        &&+{\tilde{\Omega}_{s}}(\frac{\sqrt{1}}{\sqrt{3}}|\phi_{1}\rangle+\frac{1}{\sqrt{3}}e^{i\epsilon_{2}t}|\phi_{2}\rangle
                                       +\frac{1}{\sqrt{3}}e^{i\epsilon_{3}t}|\phi_{3}\rangle)\langle\zeta|+H.c., \cr\cr
  \tilde{H}_{e}^{re}&=&\Delta|\phi_{1}\rangle\langle\phi_{1}|+\frac{\Delta}{2}(|\phi_{2}\rangle+|\phi_{3}\rangle)(\langle \phi_{2}|+\langle\phi_{3}|).
\end{eqnarray}
Consider $\epsilon_{2},\epsilon_{3}\gg\tilde{\Omega}_{1}/\sqrt{6},\tilde{\Omega}_{s}/\sqrt{3}$, we neglect terms containing high oscillating frequencies and terms
decoupled to the time evolution of initial state. Then we obtain an effective Hamiltonian
\begin{eqnarray}\label{eq2-7}
  \tilde{H}_{Z}&=&(-\frac{\sqrt{2}\tilde{\Omega}_{1}}{\sqrt{3}}|\phi_{1}\rangle\langle\psi_{1}|+\frac{\tilde{\Omega}_{s}}{\sqrt{3}}|\phi_{1}\rangle\langle\zeta|+H.c.)\cr\cr
                 &&+\Delta|\phi_{1}\rangle\langle\phi_{1}|.
\end{eqnarray}
Then by adiabatically eliminating the state $|\phi_{1}\rangle$ under large detuning condition
$\frac{\sqrt{2}\tilde{\Omega}_{1}}{\sqrt{3}},\frac{\tilde{\Omega}_{s}}{\sqrt{3}}\ll\Delta$, we obtain
an effective Hamiltonian
\begin{eqnarray}\label{eq2-8}
  H_{eff}&=&-\frac{2|\tilde{\Omega}_{1}|^{2}}{3\Delta}|\psi_{1}\rangle\langle\psi_{1}|-\frac{|\tilde{\Omega}_{s}|^{2}}{3\Delta}|\zeta\rangle\langle\zeta| \cr\cr
           &&+(\frac{\sqrt{2}\tilde{\Omega}_{1}\tilde{\Omega}_{s}^{*}}{3\Delta}|\zeta\rangle\langle\psi_{1}|
             +\frac{\sqrt{2}\tilde{\Omega}_{1}^{*}\tilde{\Omega}_{s}}{3\Delta}|\psi_{1}\rangle\langle\zeta|).
\end{eqnarray}
When we choose $\tilde{\Omega}_{s}=\tilde{\Omega}_{x}$ and
$\tilde{\Omega}_{1}=-\frac{i\tilde{\Omega}_{x}}{\sqrt{2}}$ (here
$\tilde{\Omega}_{x}$ is a real number),
\begin{eqnarray}\label{eq2-b}
  H_{eff}&=&-\tilde{\Omega}|\psi_{1}\rangle\langle\psi_{1}|-\tilde{\Omega}|\zeta\rangle\langle\zeta|+(-i\tilde{\Omega}|\zeta\rangle\langle\psi_{1}|+H.c.)\cr
         &=&-\tilde{\Omega}\cdot I+(-i\tilde{\Omega}|\zeta\rangle\langle\psi_{1}|+H.c.)
\end{eqnarray}
where $\tilde{\Omega}=\frac{{\tilde{\Omega}}_{x}^{2}}{3\Delta}$.
It is not hard to find, the first term in eq. (\ref{eq2-b}) only affects the global phase for the dynamics governed by $H_{eff}$.
So, in fact, we can directly take off the first term and further simplify the Hamiltonian as $H_{eff}=-i\tilde{\Omega}|\zeta\rangle\langle\psi_{1}|+H.c.$
when we pay no attention to the global phase. That
means, as long as $\tilde{\Omega}=\dot{\theta}$ and
$H_{eff}(t)=H(t)$; the Hamiltonian for speeding up the adiabatic
dark-state evolution governed by $H_{0}$ under condition
$\tilde{\Omega}_{1},\tilde{\Omega}_{s},\ll\sqrt{3}\lambda,\sqrt{3}\Delta$
has been constructed. As we mentioned above, to create a three-atom
$W$ state by adiabatic dark-state evolution, the Riba frequencies
$\Omega_{1}$ and $\Omega_{s}$ can be chosen in the form in eq.
(\ref{eq1a-2}). Hence, ${\Omega}_{x}$ is given
\begin{eqnarray}\label{eq2-9}
  \tilde{\Omega}_{x}=\sqrt{3\Delta\dot{\theta}}=\sqrt{\frac{\sqrt{2}\Delta(\dot{\Omega}_{1}\Omega_{s}-\dot{\Omega}_{s}\Omega_{1})}{\beta^{2}}}.
\end{eqnarray}
If we set $t'=\frac{t}{t_{f}}$, according to eq. (\ref{eq1a-2}) we can obtain two dimensionless parameters
\begin{eqnarray}\label{eq2-10}
  y_{1}&=&\frac{t't_{f}-t_{0}-0.5t_{f}}{t_{c}}, \cr\cr
  y_{2}&=&\frac{t't_{f}+t_{0}-0.5t_{f}}{t_{c}}.
\end{eqnarray}
Therefore, putting eqs. (\ref{eq1b-2}) and (\ref{eq2-10}) into eq. (\ref{eq2-9}),
we obtain
\begin{eqnarray}\label{eq2-11}
  \tilde{\Omega}_{x}=\sqrt{\frac{6\sqrt{2}\Delta G^{2}}{t_{f}}},
\end{eqnarray}
where
\begin{eqnarray}\label{eq2-12}
  G=\sqrt{-\frac{y_{1}\Omega_{1}\Omega_{2}-\Omega_{1}\Omega_{0}(y_{2}e^{-y_{2}^{2}}+\cos{\alpha}y_{1}e^{-y_{1}^{2}})}{2\Omega_{1}^{2}+\Omega_{s}^{2}}},
\end{eqnarray}
is a dimensionless wave function. A brief analysis of $G$ tells us
that the amplitude of $G$ is close to $1$. That is, the amplitude of
$\tilde{\Omega}_{x}$ is mainly dominated by
$\nu=\sqrt{\frac{6\sqrt{2}\Delta}{t_{f}}}$. According to the limited
conditions above, we have
\begin{eqnarray}\label{eq2-13}
  \sqrt{\frac{6\sqrt{2}\Delta}{t_{f}}}&\ll&\sqrt{3}{\lambda}\Rightarrow t_{f}\gg\frac{2\sqrt{2}\Delta}{\lambda^{2}}, \cr\cr
  \sqrt{\frac{6\sqrt{2}\Delta}{t_{f}}}&\ll&\sqrt{3}\Delta\Rightarrow t_{f}\gg\frac{2\sqrt{2}}{\Delta}.
\end{eqnarray}
When $\lambda$ is a constant value, under the premise that the interaction time $t_{f}$ is short, to meet the first
condition in eq. (\ref{eq2-13}), it is better to choose a smaller $\Delta$, while to meet the second condition, a larger $\Delta$ is required.
This is also demonstrated in Fig. \ref{Deltatf} which shows the
fidelity of the $W$ state in the shortcut scheme versus parameters
$\lambda t_{f}$ and $\Delta/\lambda$. We can find, too small or too
large $\Delta$ cause a long operation time for the scheme. Then, we
choose a set of suitable parameters
\{$\Delta=3\lambda,t_{f}=35/\lambda$\} for the scheme. As shown in
Fig. \ref{Ft} (a), with these parameters, we can achieve a perfect
populations transfer after very slightly correcting a related
parameter ($\tilde{\Omega}_{1}\rightarrow1.04\tilde{\Omega}_{1}$) by
numerical simulation. The parameter should be slightly corrected
because speeding up the evolution needs to slightly broke the Zeno
condition, and that cause a slight failure of the approximation. We
plot the time evolution of the populations in intermediate states
$|\psi_{3}\rangle$ and $|\phi_{1}\rangle$ in Fig. \ref{Ft} (b) to
prove this operation is necessary. In the figure, the state
$|\psi_{3}\rangle$ should have been neglected is slightly populated
(the Zeno condition is slightly broken) while the state
$|\phi_{1}\rangle$ is negligible (the second condition in eq.
(\ref{eq2-13}) is fulfilled). We give a comparison of the fidelities
via these three different methods in Fig. \ref{Ft} (c). Contrasting
with the Zeno method, the advantage of the present shortcut method
is obvious: the shortcut scheme is more robust against operational
imperfection than the Zeno one, especially, it is not necessary to
control the interaction time accurately. As demonstrated in Fig.
\ref{deltaTZ}, the fidelity almost keeps unchanging with the
variation $\delta T$, where $T=40/\lambda$ is the total operation
time chosen to complete the scheme, and a deviation
$|\delta\nu/\nu|=5\%$ which means the variation in the amplitude of
$\tilde{\Omega}_{x}$ only causes a reduction about 1\% in the
fidelity.

\begin{figure}
 \renewcommand\figurename{\small FIG.}
 \centering \vspace*{8pt} \setlength{\baselineskip}{10pt}
 \subfigure[]{
 \includegraphics[scale = 0.18]{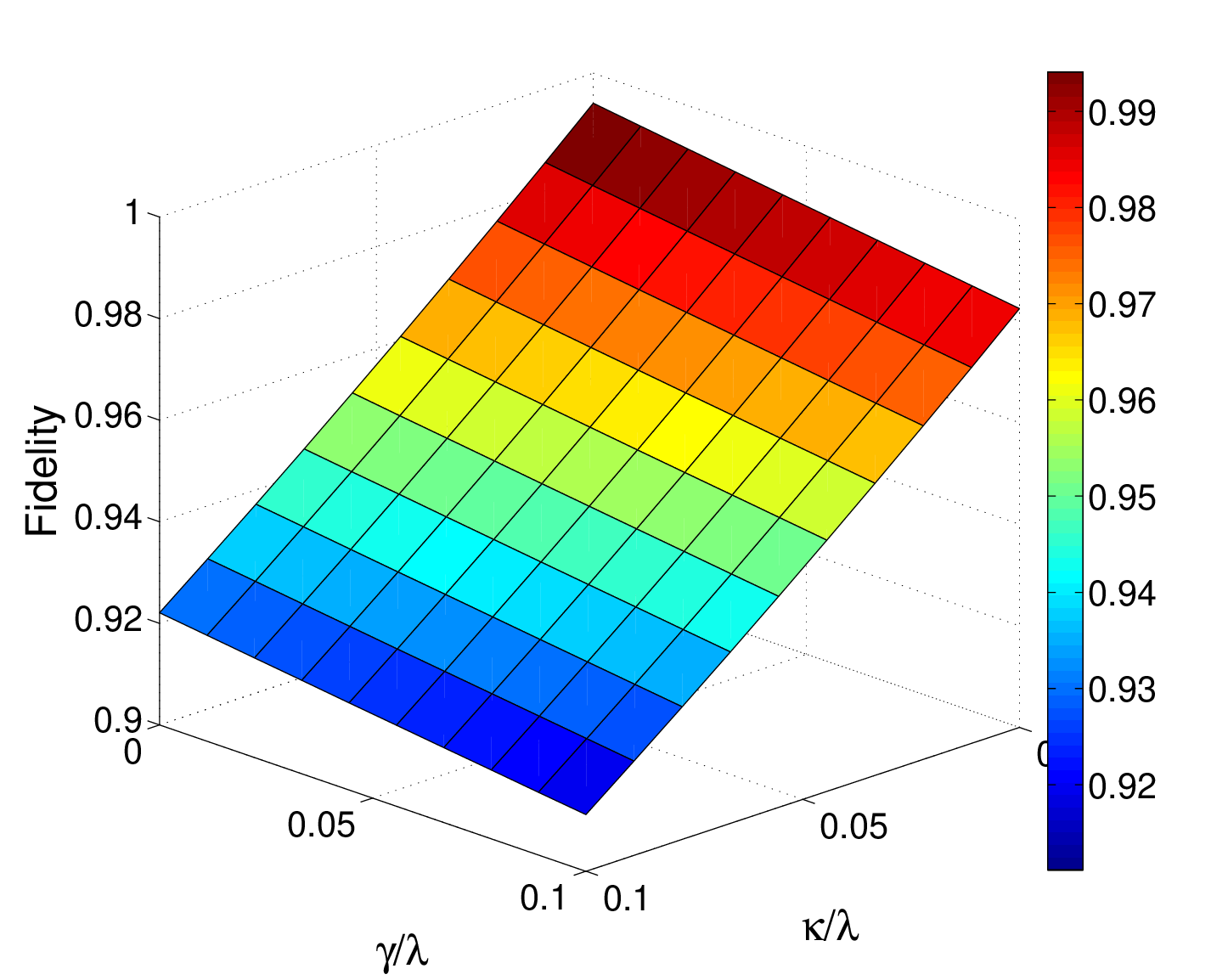}}
 \subfigure[]{
 \includegraphics[scale = 0.18]{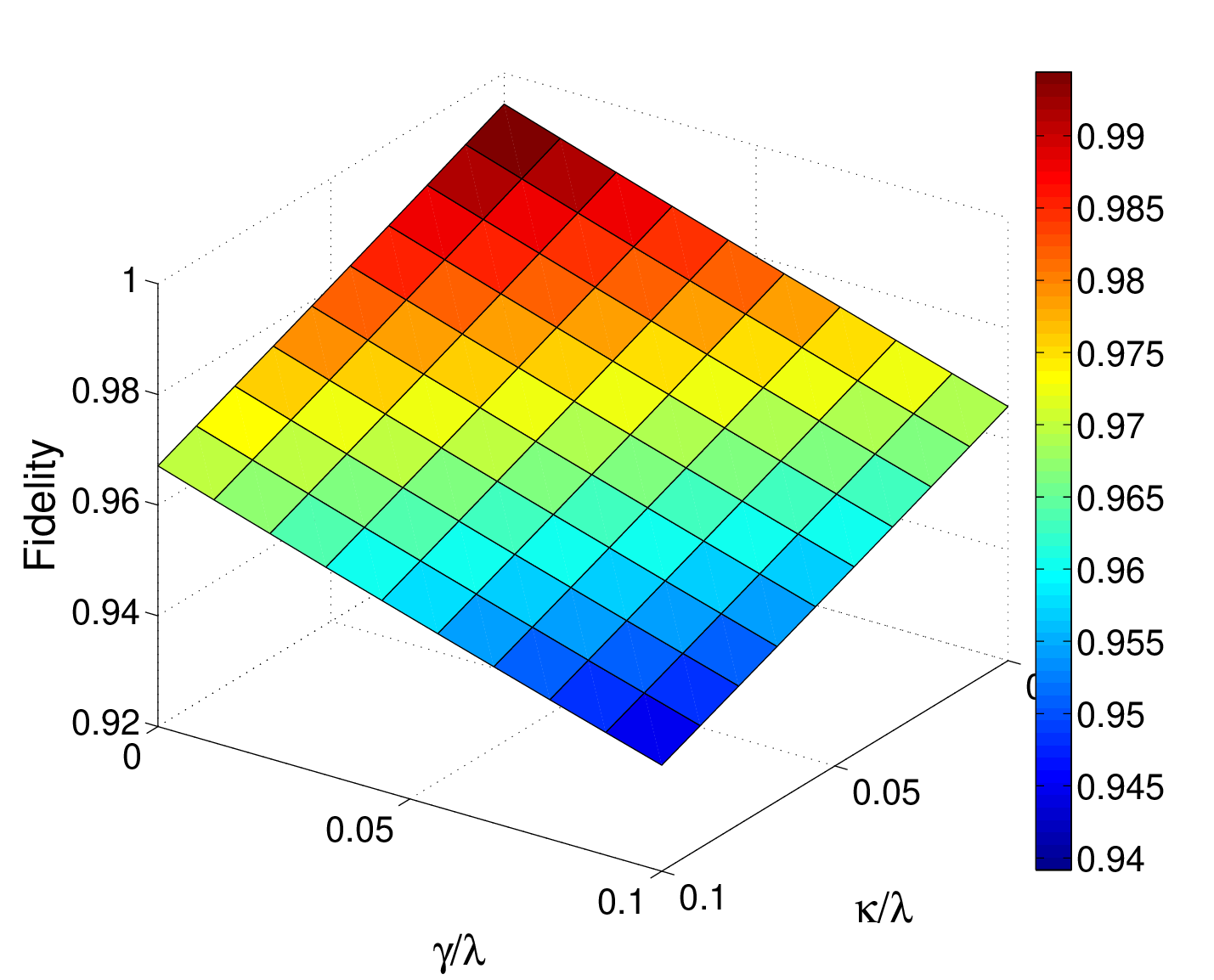}}
 \caption{Dependences on $\kappa/\lambda$ and $\gamma/\lambda$ of the fidelity of the $W$ state governed by the APF Hamiltonian
          when (a) $t_{f}=35/\lambda$ and $\Delta=3\lambda$;
               (b) $t_{f}=35/\lambda$ and $\Delta=\lambda$.}
 \label{Fkr}
\end{figure}

Now, we will check the robustness of the shortcut scheme against
possible mechanisms of decoherence. The evolution of the system can
be modeled by a master equation in Lindblad form when the
decoherence is considered,
\begin{eqnarray}\label{eq2-14}
  \dot{\rho}=i[\rho,\tilde{H}_{0}]+\sum_{k}[L_{k}\rho L_{k}^{\dag}-\frac{1}{2}(L_{k}^{\dag}L_{k}\rho+\rho L_{k}^{\dag}L_{k})],
\end{eqnarray}
where $\rho$ is the density operator for the whole system and $L_{k}$ are the Lindblad operators.
For the shortcut scheme, there are seven lindblad operators governing the dissipation:
\begin{eqnarray}\label{eq2-15}
  L_{1}^{a}&=&\sqrt{\gamma_{1}}|f\rangle_{1}\langle e|,\
  L_{2}^{a}=\sqrt{\gamma_{2}}|g\rangle_{1}\langle e|,\  \cr
  L_{3}^{a}&=&\sqrt{\gamma_{3}}|f\rangle_{2}\langle e|,\
  L_{4}^{a}=\sqrt{\gamma_{4}}|g\rangle_{2}\langle e|,\ \cr
  L_{5}^{a}&=&\sqrt{\gamma_{5}}|f\rangle_{3}\langle e|,\
  L_{6}^{a}=\sqrt{\gamma_{6}}|g\rangle_{3}\langle e|,\  \cr
  L_{7}^{c}&=&\sqrt{\kappa}a,
\end{eqnarray}
where $\gamma_{m}$ ($m=1,2,\cdots,6$) are the atomic spontaneous
emissions and $\kappa$ is the cavity decay. We set
$\gamma_{m}=\gamma/2$ for simplicity. Then by numerically solving
the master equation in eq. (\ref{eq2-15}), we plot the fidelity of
the $W$ state in the shortcut scheme versus $\gamma/\lambda$ and
$\kappa/\lambda$ in Fig. \ref{Fkr} (a). We can find that the
shortcut scheme is more sensitive to the cavity decay than atomic
spontaneous emissions with parameters \{$\Delta=3\lambda,\
t_{f}=35/\lambda$\}. The reason has been mentioned above that with
this set of parameters, the Zeno condition is not satisfied
faultlessly. So the states $|\phi_{2}\rangle$ and $|\phi_{3}\rangle$
containing cavity-excited state $|\psi_{3}\rangle$ are populated in
a certain extent during the evolution. Furthermore, the parameters
can be selected properly to restrain the cavity decay in the
experiment according to eq. (\ref{eq2-13}). For example, when we
choose \{$\Delta=\lambda$, $t_{f}=35$\}, the Zeno condition can be
satisfied well. We plot Fig. \ref{Fkr} (b) depicting the fidelity of
the $W$ state governed by the APF Hamiltonian versus
$\kappa/\lambda$ and $\gamma/\lambda$ when \{$\Delta=\lambda$,
$t_{f}=35$\}. It shows that the influence of cavity decay is
restrained with these parameters. However, it is without doubt that
the scheme is robust because the fidelity decreases slowly and even
when $\gamma=\kappa=0.1\lambda$, we still can create a $W$ state
with a high fidelity $91.12\%$.

This scheme can be easily generalized to generate $N$-atom $W$ states. We assume $N$ $\Lambda$-type atoms are trapped
in a cavity. For the original Hamiltonian, the atomic level configuration of each atom is the same as that in Fig. \ref{model} (a), and for the APF Hamiltonian,
the atomic level configuration of each atom is the same as that in Fig. \ref{model} (c).
Suppose that the $N$ atoms `see' the same field, and spatial
separation size of these atoms is much bigger than the wavelength of the emitted radiation, so
the atomic dipole-dipole interaction can be omitted. In this case, the interaction Hamiltonian for the original Hamiltonian reads
\begin{eqnarray}\label{eq2-151}
  H_{I}^{N}=\sum_{k=1}^{N}{\Omega_{k}(t)|e\rangle_{k}\langle f|+\lambda_{k}a|e\rangle_{k}\langle g|+H.c.},
\end{eqnarray}
and the APF Hamiltonian reads
\begin{eqnarray}\label{eq2-152}
  \tilde{H}_{I}^{N}=\sum_{k=1}^{N}{\Delta|e\rangle_{k}\langle e|+\tilde{\Omega}_{k}(t)|e\rangle_{k}\langle f|+\tilde{\lambda}_{k}a|e\rangle_{k}\langle g|+H.c.}.
\end{eqnarray}
We consider that the initial state of the system is in $|f,g,g,\cdots,g\rangle_{1,2,3,\cdots,N}|0\rangle_{c}$.
For the atom $1$, the classical field drives the transition resonantly between the level
$|f\rangle_{1}$ and $|e\rangle_{1}$ with the Rabi frequency $\Omega_{1}$. Then, atom $1$ will emit
a photon which will be absorbed by one of the other $N-1$ atoms with the same probability when $\lambda_{1}=\lambda_{2}=\cdots=\lambda_{N}=\lambda$.
Therefore, the excited process of $N-1$ atoms (in this part, the ``$N-1$ atoms'' means the atoms except the atom $1$) can be described by the state
$|\Psi_{e}\rangle=\frac{1}{\sqrt{N-1}}(|e,g,g,\cdots,g\rangle+|g,e,g,\cdots,g\rangle+|g,g,e,\cdots,g\rangle+\cdots+|g,g,g,\cdots,e\rangle)_{2,3,4,\cdots,N}$.
Then, by setting $\Omega_{2}=\Omega_{3}=\cdots=\Omega_{N}=\Omega_{s}$, the classical fields will drive the state $|\Psi_{e}\rangle$ to
$|\Psi_{f}\rangle=\frac{1}{\sqrt{N-1}}(|f,g,g,\cdots,g\rangle+|g,f,g,\cdots,g\rangle+|g,g,f,\cdots,g\rangle+\cdots+|g,g,g,\cdots,f\rangle)_{2,3,4,\cdots,N}$.
Hence, the single-excitation subspace could be spanned by
\begin{eqnarray}\label{eq2-153}
  |\psi_{1}\rangle&=&|f,g,g,\cdots,g\rangle_{1,2,3,\cdots,N}|0\rangle_{c}, \cr
  |\psi_{2}\rangle&=&|e,g,g,\cdots,g\rangle_{1,2,3,\cdots,N}|0\rangle_{c}, \cr
  |\psi_{3}\rangle&=&|g,g,g,\cdots,g\rangle_{1,2,3,\cdots,N}|1\rangle_{c}, \cr
  |\mu\rangle&=&|g\rangle_{1}|\Psi_{e}\rangle|0\rangle_{c}, \cr
  |\zeta\rangle&=&|g\rangle_{1}|\Psi_{f}\rangle|0\rangle_{c}.
\end{eqnarray}
Meanwhile, the Hamiltonian in the single-excitation subspace can be written as
\begin{eqnarray}\label{eq2-154}
  H_{I}^{N}&=&\Omega_{1}|\psi_{2}\rangle\langle\psi_{1}|+{\Omega_{s}}|\mu\rangle\langle\zeta|+\lambda(|\psi_{2}\rangle+\sqrt{N-1}|\mu\rangle)\langle\psi_{3}|+H.c..
\end{eqnarray}
Similarly, the APF Hamiltonian in the single-excitation subspace is
\begin{eqnarray}\label{eq2-155}
  \tilde{H}_{I}^{N}&=&\Delta(|\psi_{2}\rangle\langle\psi_{2}|+|\Psi_{e}\rangle\langle\Psi_{e}|)\cr\cr
                     &&+[\tilde{\Omega}_{1}|\psi_{2}\rangle\langle\psi_{1}|
                     +{\tilde{\Omega}_{s}}|\mu\rangle\langle\zeta|+\lambda(|\psi_{2}\rangle+\sqrt{N-1}|\mu\rangle)\langle\psi_{3}|+H.c.].
\end{eqnarray}
Obviously, the Hamiltonians in eqs. (\ref{eq2-154}) and (\ref{eq2-155}) are in the same form with those in eqs. (\ref{eq1b-1}) and (\ref{eq2-6}), respectively.
Therefore, similar as above, under the condition $\Omega_{1},\Omega_{s},\tilde{\Omega}_{1},\tilde{\Omega}_{s}\ll|\epsilon_{\pm}|$ and
$\frac{\sqrt{N-1}}{\sqrt{N}}\tilde{\Omega}_{1},\frac{1}{\sqrt{N}}\tilde{\Omega}_{s}\ll\Delta$,
where $\epsilon_{\pm}=\pm\sqrt{N}\lambda$ are the nonzero eigenvalues of $H_{ac}^{N}=\lambda(|\psi_{2}+\sqrt{N-1}|\mu\rangle)\langle\psi_{3}|+H.c.$,
$H_{I}^{N}$ and $\tilde{H}_{I}^{N}$ will be approximated as
$H_{Z}^{N}=-\frac{\sqrt{N-1}\Omega_{1}}{\sqrt{N}}|\phi_{1}\rangle\langle\psi_{1}|+\frac{\Omega_{s}}{\sqrt{N}}|\phi_{1}\rangle\langle\zeta|+H.c.$ and
$\tilde{H}_{Z}^{N}=-\frac{(N-1)|\tilde{\Omega}_{1}|^{2}}{N\Delta}|\psi_{1}\rangle\langle\psi_{1}|-\frac{|\tilde{\Omega}_{s}|^{2}}{N\Delta}|\zeta\rangle\langle\zeta|
+(\frac{\sqrt{N-1}\tilde{\Omega}_{1}\tilde{\Omega}_{s}^{*}}{N\Delta}|\zeta\rangle\langle\psi_{1}|+H.c.)$, respectively.
By setting $\tilde{\Omega}_{1}=-\frac{i\tilde{\Omega}_{x}}{\sqrt{N-1}}$ and $\tilde{\Omega}_{2}=\tilde{\Omega}_{3}=\cdots=\tilde{\Omega}_{N}=\tilde{\Omega}_{x}$,
the effective Hamiltonian which is equivalent to the counter-diabatic driving Hamiltonian of the original Hamiltonian will be achieved. Then, the shortcut can
be constructed and the $N$-qubit $W$ states can be rapidly generated.

In a real experiment, the cesium atoms which have been
cooled and trapped in a small optical cavity in the strong-coupling regime \cite{JYDWVHJKPrl99,JMJRBADBAKHCNDMSKHJKPrl03} can be used in this scheme.
On the other hand, a set of cavity QED parameters $(\lambda,\gamma,\kappa)/2\pi=(750,2.62,3.5)$MHz is predicted to be available in an optical cavity
\cite{SMSTJKKJVKWGEWHJKPra05}. With these parameters, the fidelity of the $W$ state in the shortcut scheme is $99.01\%$.

\section{conclusion}
In this paper, we have proposed a scheme to fast generate $W$ states
via transitionless-based shortcuts. In order to highlight the
advantages of the present scheme, we have described two similar
schemes based on STIRAP and QZD.
The comparison among these three schemes demonstrates that the
shortcut scheme is faster than the adiabatic one, and more robust
  against operational imperfection than the Zeno one.
Numerical investigation also demonstrates the present scheme is
robust against the decoherence caused by both atomic spontaneous
emission and photon leakage. When it comes to the generation of
$N$-atom $W$ states, the only change is setting
$\tilde{\Omega}_{1}=-\frac{i\tilde{\Omega}_{x}}{\sqrt{N-1}}$ and
$\tilde{\Omega}_{2}=\tilde{\Omega}_{3}=\cdots=\tilde{\Omega}_{N}=\tilde{\Omega}_{x}$.
Known from ref. \cite{XQSHFWLCSZYFZKHYNjp10}, the Hamiltonian for a
system with three four-level atoms trapped in a cavity also can be
approximated into an effective Hamiltonian in form of eq.
(\ref{eq1b-4}). For a similar model described in ref.
\cite{ZCSYXJSHSSQip13}, if the atomic transitions are non-resonant,
one can also obtain an effective Hamiltonian in form of eq.
(\ref{eq2-8}). That means, with the same method in section
\textrm{III}, a multi-qubit singlet state also can be fast
generated. In addition, the shortcut method might also show its glamour in other fields,
for example, fast transfer of entanglement \cite{YHCYXQQCJSLpl14,TQGJYPra14} That demonstrates the present method has a wide rang of
application in quantum information processing. This might lead to a
useful step toward realizing fast and noise-resistant quantum
information processing for multi-qubit systems in current
technology.

\section{Acknowledgments}

  This work was supported by the National Natural Science Foundation of China under Grants No.
11575045 and No. 11374054, the Foundation of Ministry of Education
of China under Grant No. 212085, and the Major State Basic Research
Development Program of China under Grant No. 2012CB921601.


\begin{thebibliography}{999}
  \bibitem{JSBPhys65} 
      J.~S.~Bell,
      {Physics (Lon Island City, NY)}
      \textbf{1}, 195 (1965).
  \bibitem{DMGMAHASAZAjp90} D.~M.~Greenberger, M.~A.~Horne, A.~Shimony, and A.~Zeilinger, {Am. J. Phys.}
      \textbf{58}, 1131 (1990).
  \bibitem{AKEPrl91} A.~K.~Ekert,
      {Phys. Rev. Lett.}
      \textbf{67}, 661 (1991).
  \bibitem{NGSMPrl97} N.~Gisin and S.~Massar,
      {Phys. Rev. Lett.}
      \textbf{79}, 2153 (1997).
  \bibitem{WDGVJICPRra00} W.~D\"{u}r, G.~Vidal, and J.~I.~Cirac,
      {Phys. Rev. A}
      \textbf{62}, 062314 (2000).
  \bibitem{CHBGBCCRJAPWKWPrl93} C.~H.~Bennett, G.~Brassard, C.~Crepeau, R.~Jozsa, A.~Peres, and W.~K.~Wootters,
      {Phys. Rev. Lett.}
      \textbf{70}, 1895 (1993).
  \bibitem{NBAPla05} N.~B.~An,
      {Phys. Lett. A}
      \textbf{344}, 77 (2005).
  \bibitem{SBZJob05} S.~B.~Zheng,
      {J. Opt. B}
      \textbf{7}, 10 (2005).
  \bibitem{JSYXHSSJpb07} J.~Song, Y.~Xia, and H.~S.~Song,
      {J.~Phys. B}
      \textbf{40}, 4503 (2007).
  \bibitem{TBCTJvZLLESGSAPrl09} T.~Bastin, C.~Thiel, J.~von~Zanthier, L.~Lamata, E.~Solano, and G.~S.~Agarwal,
      {Phys. Rev. Lett.}
      \textbf{102}, 053601 (2009).
  \bibitem{XWWGJYYHSMXQip09} X.~W.~Wang, G.~J.~Yang, Y.~H.~Su, and M.~Xie,
      {Quant. Info. Proc.}
      \textbf{8}, 431 (2009).
  \bibitem{RXCLTSPla11} R.~X.~Chen and L.~T.~Shen,
      {Phys. Lett. A}
      \textbf{375}, 3840 (2011).
  \bibitem{MLYXJSNBAOsa13} M.~Lu, Y.~Xia, J.~Song, and N.~B.~An,
      {J. Opt. Soc. Am. B}
      \textbf{30}, 2142 (2013).
  \bibitem{YHCYXJSQip13} Y.~H.~Chen, Y.~Xia, and J.~Song,
      {Quant. Info. Proc.}
      \textbf{12}, 3771 (2013).
  \bibitem{MPFBWSKBAjp97} M.~P.~Fewell, B.~W.~Shore, and K.~Bergmann,
      {Aust. J. Phys.}
      \textbf{50}, 281 (1997).
  \bibitem{KBHTBWSRmp98} K.~Bergmann, H.~Theuer, and B.~W.~Shore,
      {Rev. Mod. Phys.}
      \textbf{70}, 1003 (1998).
  \bibitem{NVVTHBWSKBArpc01} N.~V.~Vitanov, T.~Halfmann, B.~W.~Shore, and K.~Bergmann, {Annu. Rev. Phys. Chem.}
      \textbf{52}, 763 (2001).
  \bibitem{PKITMSRmp07} P.~Kr\'{a}l, I.~Thanopulos, and M.~Shapiro,
      {Rev. Mod. Phys.}
      \textbf{79}, 53 (2007).
  \bibitem{LBCWYLpl14} L.~B.~Chen and W.~Yang,
      {Laser Phys. Lett.}
      \textbf{11}, 105201 (2014).
  \bibitem{BMECGSJmp77} B.~Misra and E.~C.~G.~Sudarshan,
      {J. Math. Phys.}
      \textbf{18}, 756 (1977).
  \bibitem{WMIDJHJJBDJWPra90} W.~M.~Itano, D.~J.~Heinzen, J.~J.~Bollinger, and D.~J.~Wineland,
      {Phys. Rev. A}
      \textbf{41}, 2295 (1990).
  \bibitem{PKHWTHAZMAKPrl95} P.~Kwiat, H.~Weinfurter, T.~Herzog, A.~Zeilinger, and M.~A.~Kasevich,
      {Phys. Rev. Lett.}
      \textbf{74}, 4763 (1995).
  \bibitem{PFVGGMSPECGSPla00} P.~Facchi, V.~Gorini, G.~Marmo, S.~Pascazio, and E.~C.~G.~Sudarshan,
      {Phys. Lett. A}
      \textbf{275}, 12 (2000).
  \bibitem{PFSPPrl02} P.~Facchi and S.~Pascazio,
      {Phys. Rev. Lett.}
      \textbf{89}, 080401 (2002).
  \bibitem{XCILARDGOJGMPra10} X.~Chen, I.~Lizuain, A.~Ruschhaupt, D.~Gu\'{e}ry-Odelin, and J.~G.~Muga,
      {Phys. Rev. Lett.}
      \textbf{105}, 123003 (2010).
  \bibitem{ETSISMGMMACDGOARXCJGMAmop13} E.~Torrontegui, S.~Ib\'{a}\~{n}ez, S. Mart\'{i}nez-Garaot, M.~Modugno, A.~del~Campo, D.~Gu\'{e}-Odelin, A.~Ruschhaupt, X.~Chen, and J.~G.~Muga,
      {Adv. Atom. Mol. Opt. Phys.}
      \textbf{62}, 117 (2013).
  \bibitem{AdCPrl13} A.~del~Campo,
      {Phys. Rev. Lett.}
      \textbf{111}, 100502 (2013).
  \bibitem{SMKNPrspca10Pra11} S.~Masuda and K.~Nakamura,
      {Phys. Rev. A}
      \textbf{84}, 043434 (2011).
  \bibitem{YHCYXQQCJSPra14} Y.~H.~Chen, Y.~Xia, Q.~Q.~Chen, and J.~Song,
      {Phys. Rev. A}
      \textbf{89}, 033856 (2014).
  \bibitem{MLYXLTSJSNBAPra14} M.~Lu, Y.~Xia, L.~T.~Shen, J.~Song, and N.~B.~An,
      {Phys. Rev. A}
      \textbf{89}, 012326 (2014).
  \bibitem{MLYXLTSJSLp14} M.~Lu, Y.~Xia, L.~T.~Shen, and J.~Song,
      {Laser Phys.}
      \textbf{24}, 105201 (2014).
  \bibitem{YHCYXQQCJSLpl14} Y.~H.~Chen, Y,~Xia, Q.~Q.~Chen, and J.~Song,
      {Laser Phys. Lett.}
      \textbf{11}, 115201 (2014);
      {Phys. Rev. A}
      \textbf{91}, 012325 (2015).
  \bibitem{YHCYXQQCJSarXiv142} Y.~H.~Chen, Y,~Xia, J.~Song, and Q.~Q.~Chen,
      {Sci. Rep}
      \textbf{5}, 15616 (2015).

 \bibitem{Prl109100403} S.~Ib\'{a}\~{n}ez, X.~Chen, E.~Torrontegui, J.~G.~Muga, and A.~Ruschhaupt,
      {Phys. Rev. Lett.}
      \textbf{109}, 100403 (2012).
  \bibitem{Pra89053408} S.~Mart\'{i}nez-Garaot, E.~Torrontegui, X.~Chen, and J.~G.~Muga,
      {Phys. Rev. A}
      \textbf{89}, 053408 (2014).
  \bibitem{Njp16015025} T.~Opatrn\'{y} and K.~M{\o}lmer,
      {New J. Phys.}
      \textbf{16}, 015025 (2014).
  \bibitem{Pra90060301} H.~Saberi, T.~Opatrny, K.~M{\o}lmer, and A.~del~Campo,
      {Phys. Rev. A}
      \textbf{90}, 060301(R) (2014).
  \bibitem{Pra08743402} S.~Ib\'{a}\~{n}ez, X.~Chen, and J.~G.~Muga,
      {Phys. Rev. A}
      \textbf{87}, 043402 (2013).
  \bibitem{Pra89043408} E.~Torrontegui, S.~Mart\'{i}nez-Garaot, and J.~G.~Muga,
      {Phys. Rev. A}
      \textbf{89}, 043408 (2014).
  \bibitem{Pra8705250289063412} B.~T.~Torosov, G.~D.~Valle, and S.~Longhi,
      {Phys. Rev. A}
      \textbf{87}, 052502 (2013);
      \textbf{89}, 063412 (2014).

  \bibitem{JGMXCARDGOJpb09} J.~G.~Muga, X.~Chen, A.~Ruschhaup, and D.~Gu\'{e}ry-Odelin, {J. Phys. B}
      \textbf{42}, 241001 (2009).
  \bibitem{XCARSSADCDDOJGMPrl10} X.~Chen, A.~Ruschhaupt, S.~Schmidt, A.~del~Campo, D.~Gu\'{e}ry-Odelin, and J.~G.~Muga,
      {Phys. Rev. Lett.}
      \textbf{104}, 063002 (2010).
  \bibitem{XCJGMPra10} X.~Chen and J.~G.~Muga,
      {Phys. Rev. A}
      \textbf{82}, 053403 (2010).
  \bibitem{JFSPCGLPVNjp11} J.~F.~Schaff, P.~Capuzzi, G.~Labeyrie, and P.~Vignolo,
      {New J. Phys.}
      \textbf{13}, 113017 (2011).
  \bibitem{ETSIXCARDGOJGMPra11} E.~Torrontegui, S.~Ib\'{a}\~{n}ez, X.~Chen, A.~Ruschhaupt, D.~Gu\'{e}ry-Odelin, and J.~G.~Muga,
      {Phys. Rev. A}
      \textbf{83}, 013415 (2011).
  \bibitem{XCETDSJSLJGMPra11} X.~Chen, E.~Torrontegui, D.~Stefanatos, J.~S.~Li, and J.~G.~Muga,
      {Phys. Rev. A}
      \textbf{84}, 043415 (2011).
  \bibitem{ETXCMMSSARJGMNjp12} E.~Torrontegui, X.~Chen, M.~Modugno, S.~Schmidt, A.~Ruschhaupt, and J.~G.~Muga,
      {New J. Phys.}
      \textbf{14}, 013031 (2012).
  \bibitem{YLLAWZDWPra11} Y.~Li, L.~A.~Wu, and Z.~D.~Wang,
      {Phys. Rev. A}
      \textbf{83}, 043804 (2011).
  \bibitem{AdCPra11} A.~del~Campo,
      {Phys. Rev. A}
      \textbf{84}, 031606(R) (2011);
      {Eur. Phys. Lett.}
      \textbf{96}, 60005 (2011).
  \bibitem{ARXCDAJGMNjp12} A.~Ruschhaupt, X.~Chen, D.~Alonso, and J.~G.~Muga,
      {New J. Phys.}
      \textbf{14}, 093040 (2012).
  \bibitem{JFSXLSPVGLPra10} J.~F.~Schaff, X.~L.~Song, P.~Vignolo, and G.~Labeyrie,
      {Phys. Rev. A}
      \textbf{82}, 033430 (2010).
  \bibitem{JFSXLSPCPVGLEpl11} J.~F.~Schaff, X.~L.~Song, P.~Capuzzi, P.~Vignolo, and G.~Labeyrie,
      {Eur. Phys. Lett.}
      \textbf{93}, 23001 (2011).
  \bibitem{AWFZTRSTDOKTMHKSFSKUPPrl12} A.~Walther, F.~Ziesel, T.~Ruster, S.~T.~Dawkins, K.~Ott, M.~Hettrich, K.~Singer, F.~Schmidt-Kaler, and U.~Poschinger,
      {Phys. Rev. Lett.}
      \textbf{109}, 050502 (2012).
  \bibitem{SYTXCOl12} S.~Y.~Tseng and X.~Chen,
      {Opt. Lett.}
      \textbf{37}, 5118 (2012).
  \bibitem{XCETJGMPra10} X.~Chen, E.~Torrontegui, and J.~G.~Muga,
      {Phys. Rev. A}
      \textbf{83}, 062116 (2011).
  \bibitem{LCSYXJSJmo14} L.~C.~Song, Y.~Xia, and J.~Song,
      {J. Mod. Opt.}
      \textbf{61}, 1290 (2014).
  \bibitem{YLSLSQCWXJSZarxiv14} Y.~Liang, S.~L.~Su, Q.~C.~Wu, X.~Ji, and S.~Zhang,
      {Opt. Exp.}
      \textbf{23}, 005064 (2015).
  \bibitem{YLXJarxiv14} Y.~Liang, Q.~C.~Wu, S.~L.~Su, and S.~Zhang,
      {Phys. Rev. A}
      \textbf{91}, 032304 (2014).
  \bibitem{MVBJpa09} M.~B.~Berry,
      {J. Phys. A}
      \textbf{42}, 365303 (2009).
  \bibitem{JYDWVHJKPrl99} J.~Ye, D.~M.~Vernooy, and H.~J.~Kimble,
      {Phys. Rev. Lett.}
      \textbf{83}, 4987 (1999).
  \bibitem{JMJRBADBAKHCNDMSKHJKPrl03} J.~McKeever, J.~R.~Buck, A.~D.~Boozer, A.~Kuzmich, H.~C.~N\:{a}gerl, D.~M.~Stamper-Kurn, and H.~J.~Kimble,
      {Phys. Rev. Lett.}
      \textbf{90}, 133602 (2003).
  \bibitem{SMSTJKKJVKWGEWHJKPra05} S.~M.~Spillane, T.~J.~Kippenberg, K.~J.~Vahala, K.~W.~Goh, E.~Wilcut, and H.~J.~Kimble,
      {Phys. Rev. A}
      \textbf{71}, 013817 (2005).
  \bibitem{XQSHFWLCSZYFZKHYNjp10} X.~Q.~Shao, H.~F.~Wang, L.~Chen, S.~Zhang, Y.~F.~Zhao, and K.~H.~Yeon,
      {New J. Phys.}
      \textbf{12}, 023040 (2010).
  \bibitem{ZCSYXJSHSSQip13} Z.~C.~Shi, Y.~Xia, J.~Song, and H.~S.~Song,
      {Quant. Info. Proc.}
      \textbf{12}, 411 (2013).
  \bibitem{TQGJYPra14} T.~Qiu and G.~J.~Yang,
      {Phys. Rev. A}
      \textbf{89}, 052312 (2014).
\end{thebibliography}
\end{document}